\documentclass[12pt,preprint]{aastex}

\shorttitle{Bolocam 1.1 mm Lockman Hole Galaxy Survey}
\shortauthors{Laurent et~al.}

\begin{document}


\title{The Bolocam 1.1 mm Lockman Hole Galaxy Survey: SHARC II 350 $\mu$m Photometry and Implications for Spectral
Models, Dust Temperatures, and Redshift Estimation}



\author{
G.~T. Laurent\altaffilmark{1,2},
J. Glenn\altaffilmark{1},
E. Egami\altaffilmark{3},
G.~H. Rieke\altaffilmark{3},
R.~J. Ivison\altaffilmark{4},
M.~S. Yun\altaffilmark{5},
J.~E. Aguirre\altaffilmark{6,1},
P.~R. Maloney\altaffilmark{1}, \&
D.~Haig\altaffilmark{7}}

\altaffiltext{1}{Center for Astrophysics and Space Astronomy \& Department of Astrophysical and Planetary Sciences, University of Colorado,
    593 UCB, Boulder, CO 80309-0593}
\altaffiltext{2}{glaurent@colorado.edu}
\altaffiltext{3}{Steward Observatory, University of Arizona, 933 North Cherry Avenue, Tucson, AZ 85721}
\altaffiltext{4}{UK Astronomy Technology Centre, Royal Observatory, Blackford Hill, Edinburgh EH9}
\altaffiltext{5}{Department of Astronomy, University of Massachusetts, Amherst, MA 01002}
\altaffiltext{6}{Jansky Fellow, National Radio Astronomy Observatory}
\altaffiltext{7}{Physics and Astronomy, Cardiff University, 5 The Parade, P.O. Box 913, Cardiff CF24 3YB, Wales, UK}







\begin{abstract}

We present 350 $\mu$m photometry of all 17 galaxy candidates in the Lockman Hole detected in a 1.1 mm Bolocam survey.  Several of the galaxies were
previously detected at 850 $\mu$m, at 1.2 mm, in the infrared by {\it Spitzer}, and in the radio.  
Nine of the Bolocam galaxy candidates were detected at 350 $\mu$m and two new candidates were serendipitously detected at 350 $\mu$m (bringing
the total in the literature detected in this way to three).  Five of the galaxies have published spectroscopic redshifts, enabling investigation of the implied
temperature ranges and a comparison of photometric redshift techniques.

$\lambda$ = 350 $\mu$m lies near the spectral energy distribution peak for
$z \approx 2.5$ thermally emitting galaxies.  Thus, luminosities can be measured without extrapolating to the peak from detection wavelengths of
$\lambda \ge$ 850 $\mu$m.  
Characteristically, the galaxy luminosities lie in the range $1.0-1.2\times10^{13}$ L$_\odot$, with dust temperatures in the range of 40 K to 70 K, depending on the
choice of spectral index and wavelength of unit optical depth.  
The implied dust masses are 3 - 5 $\times$ $10^8$ M$_\odot$.  
We find that the far-infrared to radio relation for star-forming ULIRGs systematically overpredicts the radio luminosities and overestimates redshifts on the order of $\Delta z \approx 
1$, whereas redshifts based on either on submillimeter data alone or the 1.6 $\mu$m stellar bump and PAH features are more accurate.  

\end{abstract}





\section{Introduction}

Surveys at submillimeter and millimeter wavelengths have detected hundreds of galaxy candidates by their thermal dust emission.  The
galaxies (hereafter referred to as submillimeter galaxies) characteristically have redshifts $z > 1$ and inferred luminosities of $L \sim 10^{13}$ L$_\odot$ and star
formation rates of $10^3$ M$_\odot$ per year (assuming dust heating by young stars).  Such enormous luminosities and star formation rates, or accretion rates in the case
of super-massive black hole growth, make submillimeter galaxies strong candidates for the progenitors of massive galaxies at the current epoch.  Clearly, it is crucial
to characterize the spectral energy distributions (SEDs) where their emission peaks ($\lambda$ = 350 $\mu$m for 40 K dust at a redshift of $z = 2.5$), measure their
redshifts and luminosity functions, determine their power sources, and integrate them into theories of galaxy formation.

Although submillimeter galaxy SEDs peak at a few hundred microns for all but the highest redshifts, $\lambda \ge$ 850 $\mu$m surveys have been most successful at detecting
galaxies because of the lower atmospheric noise and greater transmission, and less stringent telescope surface requirements.  Most of the detections have been low
signal-to-noise ratio (just over thresholds of 3-4 $\sigma$), necessitating multiwavelength confirmation.  Furthermore, the SEDs have been extrapolated shortward from
850-1200 $\mu$m over the peak, or between 850-1200 $\mu$m and the far-infrared, to estimate dust temperatures, luminosities, and star formation rates.  
Clearly, 350 $\mu$m photometry can confirm galaxy candidates and sample the SEDs near their peaks for more precise inferences of physical parameters.  
Similarly, because of the difficulty in obtaining spectroscopic redshifts of large numbers of highly obscured galaxies, various photometric redshift estimation techniques have emerged, notably
based on the far-infrared to radio luminosity relation in ULIRGs \citep{carilli99,yun02} and the stellar continuum bump in the infrared for {\it Spitzer}-detected galaxies
\citep{egami04,sawicki02}.  However, despite the difficulty, candidate spectroscopic redshifts have been obtained for $\sim$ 73 galaxies \citep{chapman05}.  Thus, with
well-determined dust-emission SEDs, including 350 $\mu$m, photometric techniques can be compared to spectroscopic redshifts.

In this paper, we present 350 $\mu$m photometry of all 17 submillimeter galaxy candidates from the Bolocam Lockman Hole survey \citep{laurent05},
some with previous 850 $\mu$m and 1200 $\mu$m detections.  Two new galaxy candidates were serendipitously detected at 350 $\mu$m, bringing the total number of 350
$\mu$m-discovered galaxies to three \citep{khan05}.  We combine these data with infrared and radio data to derive improved luminosities, explore the range of
implied dust temperatures and spectral indices, and compare photometric redshift techniques.  This comparison is timely with the imminent launch of the {\it Herschel
Space Observatory} (scheduled for August 2007), which will detect thousands of galaxies at far-infrared and submillimeter wavelengths, but for which spectroscopic redshifts will be attainable for 
only a small fraction.  Throughout the paper, a cosmology of H$_0 = 70$ km s$^{-1}$ Mpc$^{-1}$, $\Omega_M = 0.3$, and $\Omega_\Lambda$ = 0.7 is assumed. 

\section{The 350 $\mu$m SHARC II Galaxy Survey}
\label{section:sharc}

Observations at multiple submillimeter wavelengths are vital both to confirm the Bolocam sources (as 6 false detections are 
expected from Monte-Carlo simulations) and to make photometric
redshift and temperature estimates.  The 350 $\mu$m SHARC-II observations combined with the Bolocam 1.1 mm galaxy survey provides a 
flux density ratio that is strongly dependent on redshift for a given temperature.  This is because the rest wavelength corresponding to the observed 
wavelength of 350 $\mu$m with SHARC-II is near the peak of the grey-body spectrum (for a $z \sim 2$ galaxy at 40 K), and Bolocam's 1.1 mm observations 
climb the steep $\nu^{2+\beta}$ ($\beta$ $\approx$ 1.5) modified Rayleigh-Jeans side of the spectral energy distribution.  

Follow-up observations of each of the Bolocam Lockman Hole 
galaxy candidates \citep{laurent05} were taken with the Submillimeter High Angular Resolution Camera (SHARC 
II) at the Caltech Submillimeter Observatory.  The observations were taken on three observing runs: 2004 March-April, 2005 January, and 2005 February.  The brightest Bolocam 
sources (1 and 2) were observed over 8 hours of total integration time during the 2004 March-April run, although most of the run was lost due to poor 
weather.  Bolocam sources 1 and 16 were observed over 18 hours of integration time during the 2005 January run, again with much of the run lost due 
to poor weather.  The 2005 February run was characterized by much better weather, and all the Bolocam sources except for 1, 5, 8 and 16 were 
observed over 35 hours of integration time\footnote{The primary weather measurement correlated with the SHARC II mapping speed is the CSO 225 GHz heterodyne, narrowband, 
``tipper tau'' monitor, which measures the zenith atmospheric attenuation.  The 2004 March-April, 2005 January, and 2005 February Lockman Hole 
observations yielded $\tau_{225\mathrm{GHz}}$ ranges and 75th percentiles of $\tau_{225\mathrm{GHz}}=0.046-0.087$, $\tau_{75\%}=0.076$, 
$\tau_{225\mathrm{GHz}}=0.044-0.120$, $\tau_{75\%}=0.093$ and $\tau_{225\mathrm{GHz}}=0.030-0.074$, $\tau_{75\%}=0.047$, respectively.}.  When combined with 
the observations of Bolocam sources 5 and 8 by \cite{kovacs05}, the entire Bolocam sample was observed over these observing runs.

Observations with SHARC II were taken in the point source observing mode, with a Lissajous (parametric sinusoidal curve) scan pattern using the SWEEP command of telescope.  The 
Lissajous pattern was scanned in altitude and azimuth, with amplitudes of 30$\arcsec$ and 20$\arcsec$, respectively.  When combined with the 2.6$\arcmin$ $\times$ 1.0$\arcmin$ SHARC 
II field-of-view and 9$\arcsec$ FWHM instrument beam size, this resulted in a uniform coverage region of 95$\arcsec$ $\times$ 18$\arcsec$, with a border of additional coverage (60$\arcsec$ 
$\times$ 40$\arcsec$) outside this region.  
Each observation had a fixed length of 10 minutes to ensure uniform coverage even on individual scans.  Integration times and the resulting depths of each of the SHARC II 
fields are listed in Table \ref{table:sharcdetections}.

\setlength{\tabcolsep}{1mm}
\begin{deluxetable}{ccccccccccc}
\tabletypesize{\scriptsize}
\tablecaption{SHARC II Photometry and New Galaxy Candidates}
\tablewidth{0pt}
\tablehead{
\colhead{} & \colhead{} & \colhead{} & \colhead{SHARC II/Bolocam} & \colhead{Bolocam 2 $\sigma$} & \colhead{SHARC II} & \colhead{SHARC II} & 
\colhead{} & \colhead{} & \colhead{} \\
\colhead{Bolocam} & \colhead{SHARC II} & \colhead{SHARC II} & \colhead{Offset} & \colhead{Error Circle} & \colhead{R.A} & \colhead{Dec} & 
\colhead{S/N} & \colhead{S$_\nu$} & \colhead{$\sigma$} \\
\colhead{Source} & \colhead{Source} & \colhead{\# 10 min. Scans} & \colhead{(")} & \colhead{(")} & \colhead{(J2000.0)} & \colhead{(J2000.0)} & 
\colhead{} & \colhead{(mJy)} & \colhead{(mJy)} \\
}
\startdata
1&7 &30,16&13.2		&21&10:52:57.1&57:21:01&3.7&38.0&14.0\\
2&8 &17,23&15.8		&21&10:51:18.6&57:16:36&3.5&20.9&7.9\\
3&10 &20&9.1		&21&10:52:13.0&57:15:46&3.2&14.0&5.6\\
"&11&20&17.3		&21&10:52:14.0&57:16:02&3.1&15.1&6.2\\
4&21&20&(18.3)		&22&10:52:04.8&57:18:39&&$\le$ 15.4&\\
5&6 (Kov\'{a}cs 4)&&15.0&22&10:52:30.9&57:22:06&5.9&40.4&8.6\\
6&1 & 5&12.0		&22&10:51:14.1&57:14:21&6.8&63.6&18.4\\
"&9 & 5&18.0		&22&10:51:17.8&57:14:20&3.3&27.6&10.8\\
7&20&15&(16.7)		&22&10:51:28.6&57:30:50&&$\le$ 10.7&\\
8&5 (Kov\'{a}cs 5)&&16.3&23&10:52:38.8&57:24:38&6.2&40.5&8.1\\
9&18&14&(15.4)		&23&10:53:05.0&57:15:08&&$\le$ 19.6&\\
10&16&36&(22.4)		&23&10:51:31.0&57:23:35&&$\le$ 25.1&\\
11&19&13&(19.9)		&23&10:52:49.0&57:13:01&&$\le$ 18.9&\\
12&14&18&(1.5)		&23&10:51:15.5&57:15:22&&$\le$ 20.5&\\
13&17&13&(20.5)		&23&10:52:34.9&57:18:13&&$\le$ 23.7&\\
14&2 &16&20.5		&24&10:52:01.7&57:24:43&5.9&24.1&7.3\\
15&15&15&(24.0)		&24&10:51:47.9&57:28:57&&$\le$ 20.0&\\
16&13 &93&1.6		&25&10:52:27.3&57:25:13&3.0&44.0&18.3\\
17&12&16&6.0		&26&10:52:00.6&57:24:21&3.1&15.5&6.3\\
\hline
(9)&3 &14&-		&23&10:53:08.3&57:15:01&4.8&28.4&9.2\\
(16)&4 &93&-		&25&10:52:32.3&57:24:48&3.8&37.0&13.4\\
\enddata
\tablecomments{SHARC II detections and 3 $\sigma$ upper limits at each of the Bolocam sources, in order of descending brightness at 1.1 mm.  The Bolocam sources in parentheses 
correspond to SHARC II detections well outside of the Bolocam 2 $\sigma$ positional error circle, and are therefore believed not to be associated with the Bolocam source.  Two 
SHARC II detections from \cite{kovacs05} are also included.  The SHARC II source numbers are listed in order of 350 $\mu$m S/N.  
}
\label{table:sharcdetections}
\end{deluxetable}
\setlength{\tabcolsep}{2mm}

The reduction of the raw SHARC II data was accomplished with the use of the "deep" cleaning utility of the Comprehensive Reduction Utility for SHARC 
II (CRUSH\footnote{http://www.submm.caltech.edu/$\sim$sharc/crush/index.htm}).  Observations of pointlike galaxies, quasars, protostellar sources, H$_\mathrm{II}$ regions, 
and evolved stars were used to construct pointing models 
for each of the observing runs.  Observations of the pointing sources were taken with a scan strategy identical to that of the science fields.  
A subset of the pointing sources were used for flux density calibration, with reference 350 $\mu$m flux densities obtained from the SHARC II 
website\footnote{http://www.submm.caltech.edu/$\sim$sharc/}.  

Source extraction was performed on the CRUSH-cleaned maps, with each map (corresponding to a single Bolocam candidate) consisting of all of the 
individual scans co-added together.  The algorithm was begun by doing a cut on the uniform coverage region, defined as the set of pixels for which the 
coverage is $\ge$ 60\% of the maximum per-pixel coverage.  The uniform coverage region is a contiguous region in the center of each map.  Next, an RMS in 
sensitivity units (the flux density of each pixel times the square root of the integration time for that pixel in units of mJy s$^{1/2}$) was computed in 
the uniform coverage region.  This RMS is valid for the entire uniform coverage region since variations in coverage have been
accounted for by the $t_i^{1/2}$ coverage normalization, where $t_i$ is the total integration time for pixel $i$.  All
pixels with coverage-normalized flux densities exceeding 3 $\sigma$ (``hot pixels'')
were flagged as potential sources.  Then hot pixels were grouped into multi-pixel sources by making the maximal group of adjacent hot pixels, including 
those within $\sqrt{2}$ pixels (i.e., diagonally adjacent).  The peak flux density, right ascension and declination of the source candidates were computed
by centroiding two-dimensional Gaussians on the groups.  The uncertainty in the flux density of each source is given by the 
pixel-to-pixel RMS at the centroid location of the source.

\section{Positional Uncertainties}
\label{section:positionalerrors}

The large beam sizes of submillimeter and millimeter wave instruments (31$\arcsec$, 14$\arcsec$, 11$\arcsec$, and 9$\arcsec$ FWHM for Bolocam, SCUBA, MAMBO, and SHARC II, respectively) 
makes it difficult to identify likely optical and radio counterparts to the galaxy candidates.  Despite the large beam sizes, however, individual sources 
can be centroided to much higher precision than the quoted beam size.  To help constrain this issue of source matching between the various surveys, a 
positional error circle was estimated for each of the submillimeter and millimeter band detections.  For the Bolocam detections, Monte Carlo simulations 
were performed by injecting sources into the timestream and running the reduction pipeline and source extraction algorithm.  This simulation was repeated 
for a range of source flux densities.  The resulting centroiding error as a function of flux density (5.4 - 9.1$\arcsec$) was added in quadrature with the RMS telescope 
pointing error (9.1$\arcsec$) to yield a range of 2 $\sigma$ positional error circles of 21 - 26$\arcsec$.  A similar approach was used to estimate the centroiding errors for 
SCUBA \citep{scott02} and MAMBO \citep{greve04}, yielding 2.2 - 10.4$\arcsec$ and 1.2 - 4.3$\arcsec$ respectively.  When added in quadrature to the quoted pointing errors (4$\arcsec$ and 
3$\arcsec$, respectively), this yields 2 $\sigma$ positional errors of 9.2 - 22$\arcsec$ and 6.5 - 13$\arcsec$, respectively.  As the centroiding error as a function of flux density for 
the SHARC II 
observations was not 
available, values of 3.0 - 4.0$\arcsec$ were empirically determined from the SHARC II centroiding fits of the Bolocam sources, which, when added in quadrature with the pointing error of 
3.8$\arcsec$,
yields 2 $\sigma$ positional errors of 9.8 - 11$\arcsec$.  These error circles were used to correlate the sources between the 
different surveys to find coincident detections.

\section{Results}
\subsection{SHARC II 350 $\mu$m Detections}
\begin{figure}
\epsscale{1.0}
\plotone{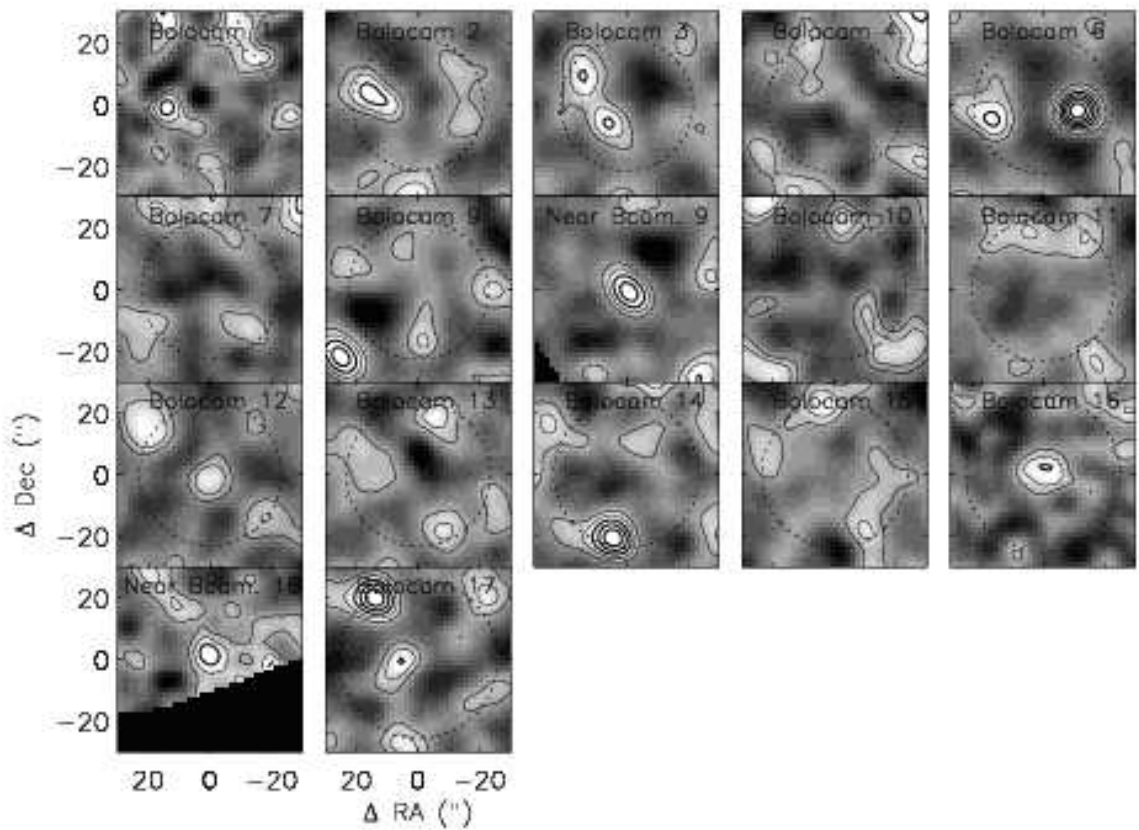}
\caption{Postage stamp images (in sensitivity units of mJy s$^{1/2}$) of each SHARC II field, centered on the Bolocam source positions. Images have been cropped to 60$\arcsec$ $\times$ 
60$\arcsec$.  The 
dotted circles represent the 2 $\sigma$ positional error circles of each Bolocam galaxy candidate.  Overlayed are contours of signal-to-noise, with levels of 1, 2, 3, 4, 5, and 6.  
Contours of positive SHARC II detections (S/N $\ge$ 3) are thicker than the others for clarity.  Bolocam sources 5 and 8 are not included, as they were observed by \cite{kovacs05}.}
\label{figure:postage}
\end{figure}

Postage stamp images of each of the SHARC II fields are shown in Figure \ref{figure:postage}.  Each image has been cropped to 60$\arcsec$ $\times$ 60$\arcsec$, 
centered on the Bolocam source positions (dotted circle).  The SHARC II source candidate list is presented in Table \ref{table:sharcdetections}, where the sources are 
listed in order of Bolocam source number.  Seven Bolocam galaxy candidates were detected by SHARC II at $> 3 \sigma$ (Bolocam 1, 2, 3, 6, 
14, 16, 17).  Two of these sources (Bolocam 3, 6) were found to have two SHARC II counterparts.  Two Bolocam candidates (5 and 8) were observed by \cite{kovacs05}
and are also included in the list.  
3 $\sigma$ upper limits are given for each of the Bolocam fields with no positive detections.  Note that the flux density uncertainties in the last column of 
Table \ref{table:sharcdetections} include uncertainties of $\sim$ 20\% due to calibration error (as determined by the dispersion of the calibration 
source flux densities).  The correlation between the two source lists based upon the positional error circles and detection offsets is 
lower than expected, as only 2 of the 7 Bolocam sources with a single SHARC II counterpart have errors within 1 $\sigma$ (5 are expected from a normal 
distribution).  This may be due to underestimating the Bolocam pointing error (cf.\ \S\ \ref{section:positionalerrors} and Laurent et al.\ 2005).

An additional two sources were detected in the survey (in the fields of Bolocam 9 and 16); they are not associated with the Bolocam sources because their locations are well outside of the 
Bolocam positional error circles (see \S\ \ref{section:positionalerrors}).  While 0.8 false detections are expected from Gaussian statistics in the SHARC II survey (given the 60\% uniform 
coverage cut and the 9\arcsec\ SHARC II beam), these serendipitous sources nevertheless may be real.  SHARC II sources 3 and 4 were detected at high significance at 350 $\mu$m and lie in 
regions of positive flux density in the Bolocam map, with S/N ratios of 2.9 and 2.1, respectively.

\subsection{SHARC II / Bolocam Correspondence}
\label{section:summary}

\begin{figure}
\epsscale{1.0}
\plotone{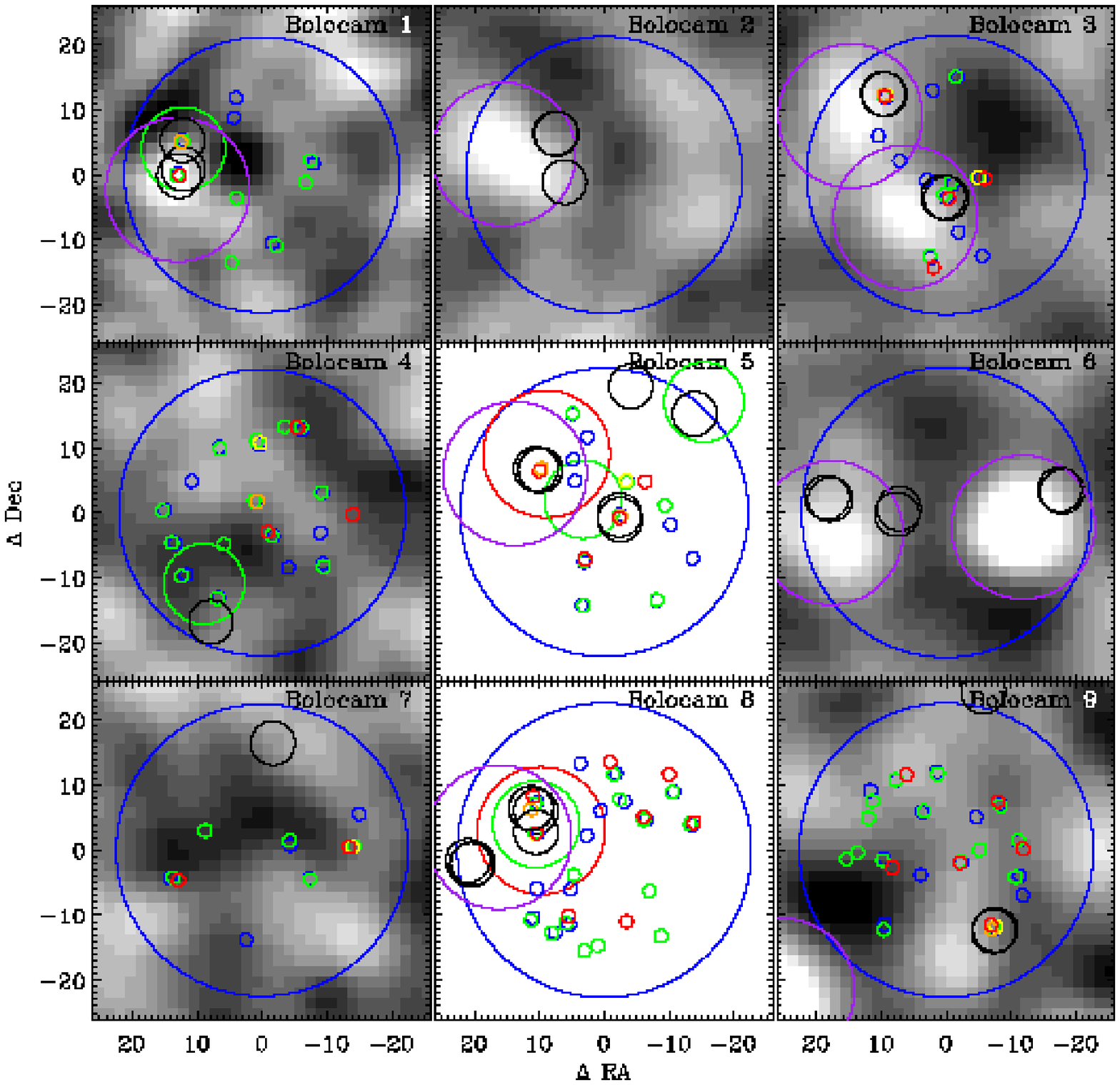}
\caption{Plot of coincident multiwavelength detections of all 17 Bolocam sources, centered on the Bolocam positions.  Sizes of the Bolocam, SCUBA, MAMBO, and SHARC
II circles correspond to the 2 $\sigma$ positional error circles.  Black circles correspond to VLA radio detections by \cite{yun05}, \cite{ivison02},
\cite{ciliegi03}, and \cite{biggs06}.  Smaller
diameter circles correspond to the {\it Spitzer} 2.6, 4.5, 5.8, 8.0 IRAC and 24 $\mu$m MIPS bands.}
\label{fig:circles}
\end{figure}

In addition to the Bolocam 1.1 mm and SHARC II 350 $\mu$m detections, existing multiwavelength coverage (submillimeter,
radio, infrared, optical, and X-ray) of the Lockman Hole was used to identify likely counterparts and characterize coincident sources.
A detailed description of each of these surveys is found in Appendix \ref{section:coverage}.  A comprehensive summary of the counterparts to the Bolocam detections (including the coverage by each
survey) is listed in Table \ref{table:detections}.  Given the large size of the Bolocam beam, identifying likely counterparts requires a certain amount of judgement.  Detailed maps of the sources
can be found in Figure \ref{fig:circles}.  Additional notes on individual objects are discussed below.

\begin{figure}
\epsscale{1.0}
\plotone{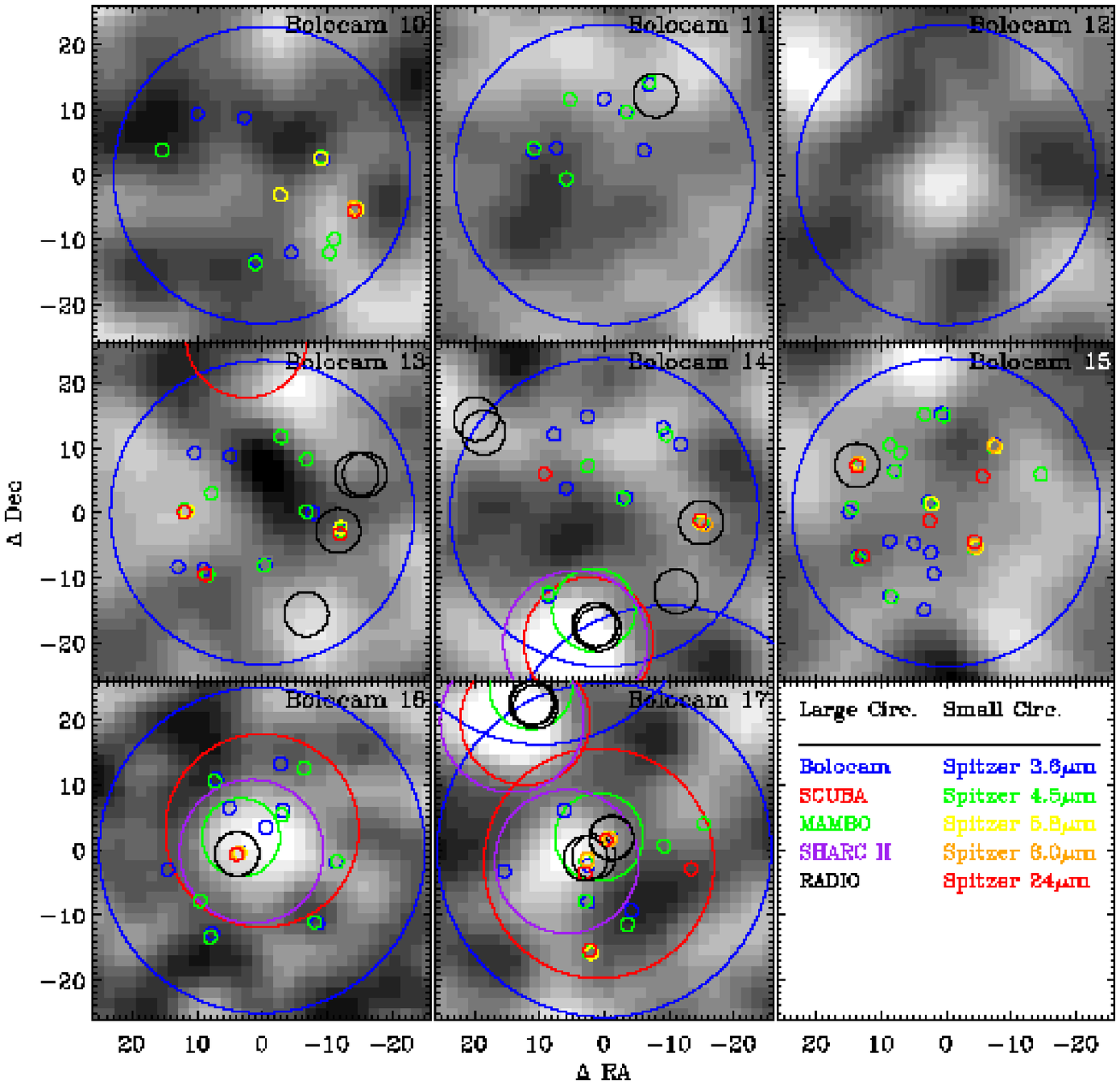}
\end{figure}

Bolocam.LE.1100.1 --
We conclude that Bolocam source 1 is likely to be a submillimeter galaxy given the coincident Bolocam, MAMBO, and SHARC II detections.  In addition, highly
plausible radio, {\it Spitzer}, and faint optical counterparts exist.

Bolocam.LE.1100.2 -- The SHARC II detection falls 16$\arcsec$ to the ENE, but contains both radio sources in its error circle.  The two 20 cm radio sources
\citep{yun05,biggs06} are unrelated due to their separation.  We treat each of the radio sources separately when fitting photometric redshifts.  The coincident northern radio source with
bright optical counterparts has a low photometric redshift (see next section) and is likely to be a low-redshift galaxy as the SDSS survey concludes.  Bolocam and
SHARC II may also be detecting the \cite{biggs06} southern radio source, which has a very faint optical counterpart.

Bolocam.LE.1100.3 -- The position of this 6.0 mJy Bolocam source is at the edge of the good coverage region of the SCUBA survey.  All of the optical
counterparts to the radio sources are relatively bright (22 - 24 magnitude), although curiously, the SDSS catalog classifies the optical counterpart to the northeast radio source as a star.  (Given the
radio and 350 $\mu$m SHARC II counterparts, we conclude that the SDSS classification may be incorrect.)  Given that both of the SHARC II sources have radio, {\it Spitzer}, and optical
counterparts, each (or both) are likely candidates as submillimeter galaxies and each could contribute to the flux density of the Bolocam source.  Note that when estimating photometric redshifts
where source confusion may be present (for this Bolocam source and elsewhere), no attempt was made to partition the Bolocam flux density among multiple submillimeter sources (due to the large
uncertainties in position).

Bolocam.LE.1100.5 -- Four radio detections fall within the Bolocam positional error circle: one within the SCUBA and SHARC II error circles and just outside the
edge of the MAMBO error circle, one on the edge of the MAMBO error circle, and the other two near the edge of the Bolocam positional error circle.
The southwest radio source has 5-color SDSS photometry and is classified as a galaxy (extended).  The fact that three of the radio sources lie outside both the SCUBA and SHARC II error circles
makes them unlikely to be the correct counterpart of the submillimeter detections.  We therefore choose the northeast radio source to be the more likely counterpart
(which is confirmed by the fact that Chapman et al.\ 2005 were able to obtain a spectroscopic reshift for this submillimeter galaxy at
this radio position, as discussed in Appendix \ref{section:previousredshift}).

\noindent A MAMBO detection located just outside the Bolocam error circle was not detected by SHARC II or SCUBA (3 $\sigma$ upper limit of 35.4
mJy), but has a \cite{chapman05} spectroscopic redshift of 1.956.  Due to the large size of the Bolocam beam, the Bolocam flux densities and positions may be influenced by source confusion.

Bolocam.LE.1100.6 -- Two SHARC II counterparts fall within the Bolocam positional error circle, each with radio counterparts (with the eastern source containing two radio counterparts).
The SDSS survey classifies both the radio source associated with the western SHARC II source and the eastern radio source associated with the eastern SHARC II source as galaxies.
Each of the three radio sources may be contributing to the Bolocam flux density due to source confusion.  We treat each of the radio sources separately when fitting photometric redshifts.

Bolocam.LE.1100.8 -- Given the fact that the southern radio source (of the pair of two radio sources oriented N-S) has Bolocam, SCUBA, MAMBO, SHARC II, and {\it
Spitzer} detections, along with a spectroscopic redshift, we conclude that this Bolocam source is real.  Nevertheless, the northern radio
source cannot be ruled out as a galaxy also contributing to the submillimeter fluxes.

Bolocam.LE.1100.14 -- This 4.4 mJy Bolocam detection is likely influenced by source confusion, given three closely spaced submillimeter sources (SCUBA sources
1, 4, and 8, with the latter two lying
near, but outside of the Bolocam positional error circle).  The location of the southern radio source relative to the Bolocam position is greater than the 16$\arcsec$ radius used for the 
{\it Spitzer} 
counterparts catalog, and thus no {\it Spitzer} data is available.  This set of coincident sources is the most likely counterpart to the Bolocam
source.  Just outside of the Bolocam error circle lies another 850 $\mu$m SCUBA detection (LE850.4) to the northeast, with a coincident \cite{ivison02} 20 cm
radio source as well as published 3.6, 4.5, 5.8, and 8.0 $\mu$m {\it Spitzer} counterparts.

The SCUBA source that coincides with Bolocam source 14 (LH850.1) is also the brightest SCUBA source and has been the target of many published multi-wavelength studies.  In addition to the
extensive radio, infrared, optical and X-ray surveys discussed in Appendix \ref{section:coverage}, a faint (K $\simeq$ 23.5) galaxy counterpart was positively identified \citep{lutz01} at
the radio position.  The source was found to be extended (20-30 kpc), clumpy (on subarcsecond scales) and very red ({\it I - K} $>$ 6.2).

Bolocam.LE.1100.16 -- We conclude that this Bolocam source is in fact a submillimeter galaxy, given the large number of multiwavelength detections and a radio source with a confirmed
spectroscopic redshift.

Bolocam.LE.1100.17 -- This 4.0 mJy Bolocam detection is likely influenced by source confusion, given two nearby submillimeter sources.  The 850 $\mu$m SCUBA,
1.1 mm MAMBO, and 850 $\mu$m SHARC II coincident detections to the northeast are the likely counterparts to Bolocam source 14 and are 
discussed in detail in Bolocam.LE.1100.14.  The SDSS catalog curiously classifies the northwest radio source as a star.  We conclude that both the southeast and northwest radio sources associated
with the Bolocam source may be submillimeter galaxies, given the large number of multiwavelength detections.  While a confirmed spectroscopic redshift exists near the southeast radio source,
self-consistent photometric redshifts and multiple optical counterparts at the radio position suggest that the spectroscopic redshift may not correspond to the radio / submillimeter sources (see
\S\ \ref{subsection:bolocam17}).

\subsection{SHARC II Non-Detections}

From extensive Monte-Carlo simulations of the Bolocam data set \citep{laurent05}, 6 false detections (Poisson distributed) are expected.  This represents a large fraction (6/17) of 
the overall source catalog and is a consequence of the relatively low 3 $\sigma$ detection threshold used in the source detection algorithm.  Eight of the Bolocam sources (4, 7, 9, 10,
11, 12, 13, and 15) were found to show no secure counterparts at 350 $\mu$m, although two of the sources (Bolocam 9, 12) exhibit flux densities just below the 3 $\sigma$ detection
threshold (with a coincident radio detection for Bolocam source 9).   Here we describe each of the SHARC II non-detections of the Bolocam sources in detail.

Bolocam.LE.1100.4 --  A single radio counterpart \citep{biggs06} lies near the edge of the MAMBO positional error circle, with an SDSS classification of the optical counterpart as a
galaxy.  While well within the Bolocam positional error circle, the location of the radio source
relative to the Bolocam position is greater than the 16$\arcsec$ radius used for the {\it Spitzer} counterparts catalog, and thus no {\it Spitzer} data is available.  Given the coincident 
Bolocam and Mambo 
detections, along with a lack of SHARC II and SCUBA detections, this source could possibly be a very high redshift galaxy ($z > 4$), such that the SED falls below the 3 $\sigma$
detection threshold of the SCUBA 850 $\mu$m survey.

Bolocam.LE.1100.7 -- The lack of multiwavelength observations makes it difficult to determine whether this Bolocam detection is real (and associated with the coincident radio detection).
The lack of {\it Spitzer} and SHARC II counterparts to the radio source, however, leads us to believe that the Bolocam source may be a spurious detection.

Bolocam.LE.1100.9 -- It is interesting to point out that the SHARC II upper limit in the Bolocam error circle is just below the
3 $\sigma$ detection flux density threshold and coincides with the radio position.  The lack of multiwavelength observations makes it difficult to determine whether this Bolocam detection is real.
The presence of {\it Spitzer} counterparts and a possible dim SHARC II detection, however, leads us to believe that the Bolocam source may be real.

Bolocam.LE.1100.10 -- Given the lack of counterparts, we conclude that there is little evidence to suggest that this detection represents a submillimeter galaxy and is likely a
spurious detection.

Bolocam.LE.1100.11 -- The lack of more multiwavelength data makes it difficult to determine whether this Bolocam detection is real.
While lacking a SHARC II detection, the radio source with an SDSS classification as a galaxy (extended object) leads us to believe that the Bolocam source
may be real.

Bolocam.LE.1100.12 -- A portion of the Bolocam error circle lies outside the deep \cite{ivison05} optical Subaru R-band field.  Similar to Bolocam source 9, we point out that the
SHARC II upper limit in the Bolocam error circle is just below the 3 $\sigma$ detection flux density threshold.  The lack of more multiwavelength data makes it
difficult to confirm whether this Bolocam detection is real.

Bolocam.LE.1100.13 -- This Bolocam source lacks 850 $\mu$m SCUBA detections (although a SCUBA source is located just
outside of the Bolocam positional error circle).  We suggest that there is little evidence that the Bolocam detection represents a submillimeter galaxy and is likely a spurious
detection.

Bolocam.LE.1100.15 --
The lack of MAMBO and SHARC II counterparts makes it difficult to confirm the Bolocam detection.  Nevertheless, the SDSS classification of the position coincident with the radio source
as a galaxy (extended) leads us to believe that the Bolocam source may be real.

\subsection{Submillimeter Spectral Energy Distributions}
\label{subsection:correlations_between_spectra}

\begin{figure}
\epsscale{1.0}
\plotone{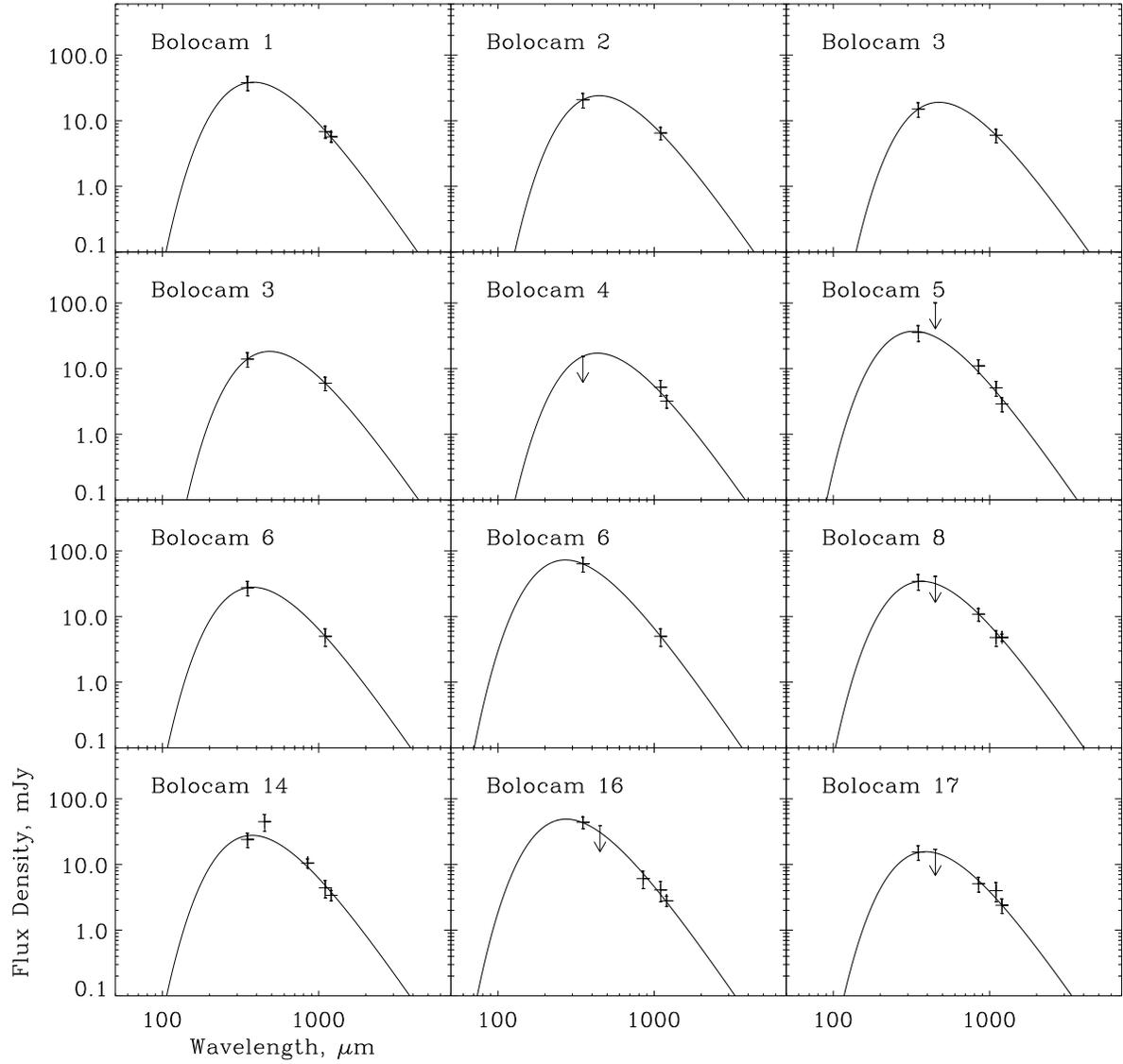}
\caption{Submillimeter SEDs for all Bolocam galaxies detected in at least one other submillimeter waveband.  The solid line represents a fit to a modified blackbody
spectrum using the parameters of Laurent et al.\ (2005, See \S\ \ref{section:photometric_redshifts}).}
\label{figure:sed_all}
\end{figure}

\begin{figure}
\epsscale{1.0}
\plotone{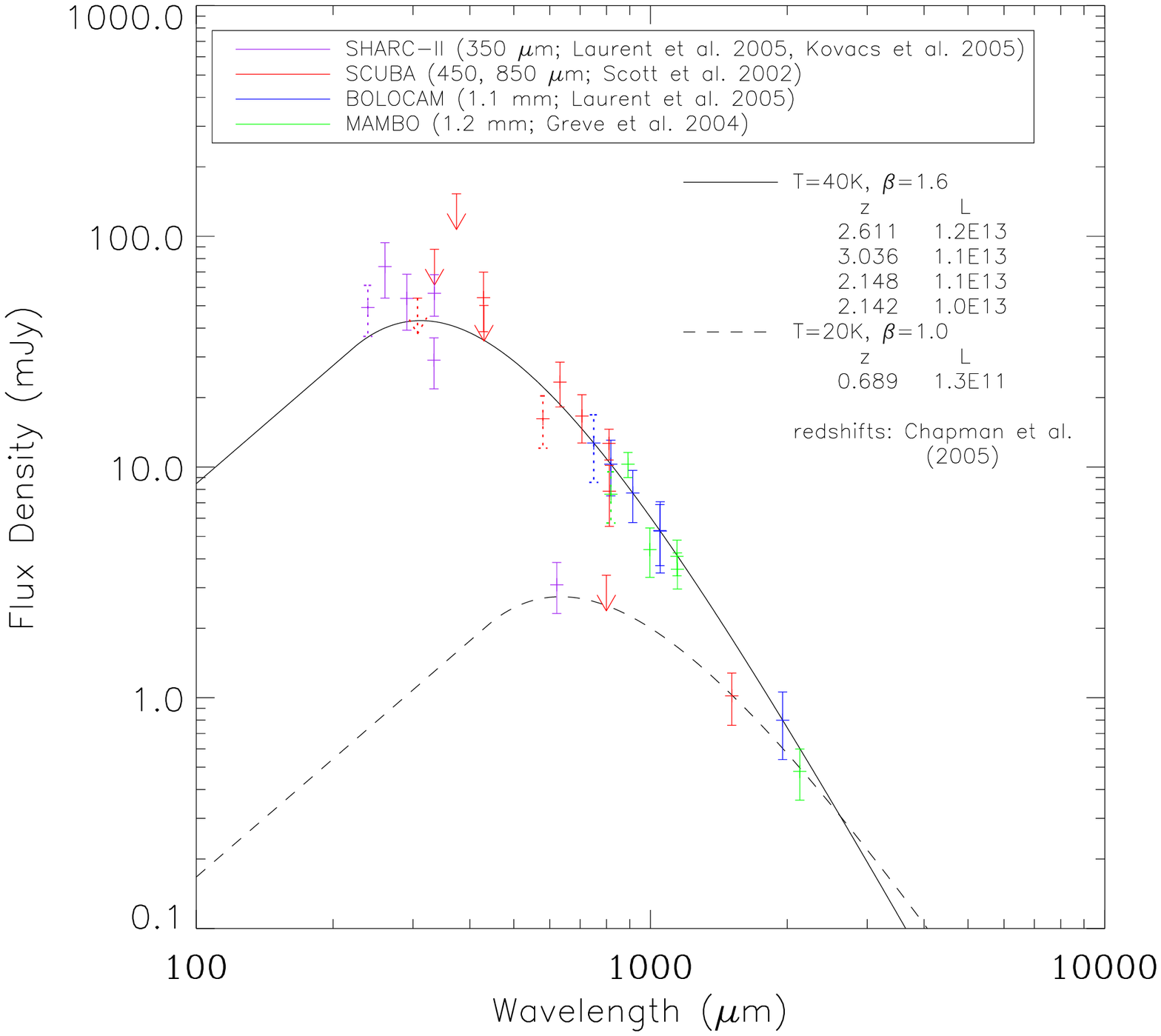}
\caption{Composite submillimeter SED for the five Bolocam galaxies with spectroscopic redshifts (Bolocam 5, 8, 14, 16, and 17).  The observed SEDs have all been redshifted to $z$ = 2.0 using the
spectroscopic redshifts, corrected for cosmological dimming, and their Bolocam 1.1 mm fluxes normalized to the \cite{laurent05} model.  The dotted points represent Bolocam source 17
assuming a redshift of 2.4 (the median photometric redshifts from \S\ \ref{section:redshift_techniques}).  The solid line represents the \cite{laurent05} model (T = 40 K, $\beta$ =
1.6) based on the composite SED of Blain et al.\ (2002, and references therein).  The dashed line represents the same model, with T = 20 K and $\beta$ = 1.0.}
\label{figure:revertzoom}
\end{figure}

The submillimeter spectral energy distributions (SEDs) of the coincident SHARC / Bolocam detections is shown in Figure \ref{figure:sed_all}.  Five of the 17 Bolocam galaxy candidates
(5, 8, 14, 16, 17) have spectroscopic redshifts from Chapman et al. (2005, see Appendix \ref{section:previousredshift}).  In order to properly compare the SEDs, it is necessary to shift each of the SEDs to a
common redshift.  Thus, each observed SED was brought to a redshift of 2.0 (the mean redshift of the five Bolocam galaxies) using the spectroscopic redshifts.  The composite SED of these five Bolocam
galaxies can be seen in Figure \ref{figure:revertzoom}.

In addition to redshifting the SEDs to align their rest wavelengths, a cosmological dimming term was applied by assuming a flat
($\Omega_k=0$), $\Omega_\Lambda$ = 0.7 cosmology.
Finally, to account for variations in their intrinsic brightnesses, we normalize the flux densities of these five Bolocam galaxies by tying
together their SEDs at the observed Bolocam wavelength of 1.1 mm.  To account for the spread of the redshifted wavelengths of the 1.1 mm
Bolocam observations, the flux densities were normalized to the \cite{laurent05} model based on the observations cited in the \cite{blain02}
paper.  The model assumes a single dust temperature of 40 K ($\beta=1.6$) and is overplotted as a solid line in
Figure \ref{figure:revertzoom}.  Note that only the Bolocam observations are constrained to pass through this model.

Upon inspection, we find that at least four of the five Bolocam galaxies with spectroscopic redshifts (5, 8, 14, 16) exhibit very similar
SEDs in the submillimeter and millimeter wavelengths.  They are modeled adequately by the 40 K composite SED based on nearby dusty IRAS
galaxies, high redshift submillimeter galaxies, gravitationally lensed high-redshift galaxies, and high redshift AGN.  The Bolocam
galaxy 17, however, appears to peak at a much higher wavelength and lower flux density than the others.  We believe that there is enough
source confusion to question whether the \cite{chapman05} redshift for this galaxy ($z$ = 0.689) corresponds to the SED shown (see \S\ \ref{subsection:bolocam17}).  If the
spectroscopic redshift is valid, the SED is modeled much better by a T = 20 K ($\beta=1.0$) grey-body dust spectrum.

\section{Redshifts}
\label{section:photometric_redshifts}

\subsection{Introduction}

With the multiwavelength photometry of the Bolocam sources, we fit photometric redshifts using various models based on 
different portions of the SED.  Photometric redshifts based on the far-IR-to-radio correlation were derived using the models 
of \cite{carilli99} and \cite{yun02}.  The shape of the submillimeter and millimeter part of the spectrum was also fit  
without the radio points, assuming a blackbody emission spectrum modified by a dust emissivity term \citep{wiklind03,laurent05}.  A brief description of each of the models, fitting 
methods, and the redshift results are discussed in the next section.

\subsection{Redshift Techniques}
\label{section:redshift_techniques}

This section attempts to briefly describe each of the five photometric redshift techniques used in this paper and the results of the fits when applied to the Bolocam galaxy candidates.  The 
following section will compare the relative merits of each of the photometric redshift fitting techniques and discuss the results of redshift distributions.

1) FIR-Radio Spectral Index -- \cite{carilli99} used the semianalytic, linear relationships 
derived by Condon (1992) between the massive star formation rate and the radio synchrotron luminosity and far-IR dust emission from active star-forming galaxies to 
show that the spectral index between these two frequencies, $\alpha^{350}_{1.4}$, is a well behaved function of redshift:
\begin{eqnarray}
\alpha^{350}_{1.4} = -0.24 - [0.42 \times (\alpha_{\mathrm{radio}} - \alpha_{\mathrm{submm}}) \times \mathrm{log} (1+z)],
\end{eqnarray}
where we adopt the standard value in Condon (1992) of -0.8 for $\alpha_{\mathrm{radio}}$, and a value of +3.2 for $\alpha_{\mathrm{submm}}$ (an average of the spectral indices between 270 
and 850 GHz for 
M82 and Arp 220.  The relation is believed to be a result of relativistic electrons accelerated in supernova remnants (producing synchrotron radiation) and dust heated by the interstellar 
radiation field (with a thermal peak of $\sim 380 \mu$m for a galaxy with $z = 2$ and T = 40 K).  Photometric redshifts determined using only the Bolocam and radio flux densities are listed in Table 
\ref{table:redshifts_yunandcarilli}.  Redshift results from Bolocam sources with multiple radio counterparts are listed using the higher S/N detection 
in the case of coincident detections by independent surveys or are listed together in the case of multiple counterparts detected by a single group.  

The error bars listed in Table \ref{table:redshifts_yunandcarilli} (and elsewhere throughout this paper) were obtained from Monte-Carlo simulations of the 
fits and represent statistical errors due to measurement uncertainty in the flux densities.  The flux densities at each observed wavelength were varied 
about their mean value assuming a Gaussian distribution of flux errors.  Each Monte-Carlo SED was 
then fit to the photometric redshift models with a standard, least-squares minimization fitting routine.  Each simulation was repeated 1000 times, with the 
error bars quoted being the minimum-length 1 $\sigma$ confidence intervals from the resulting histogram of redshifts.  It should be noted that these 
confidence intervals represent only the statistical goodness of fit and that uncertainties in the templates themselves are expected to dominate 
the photometric redshift errors.

\begin{figure}
\epsscale{1.0}
\plotone{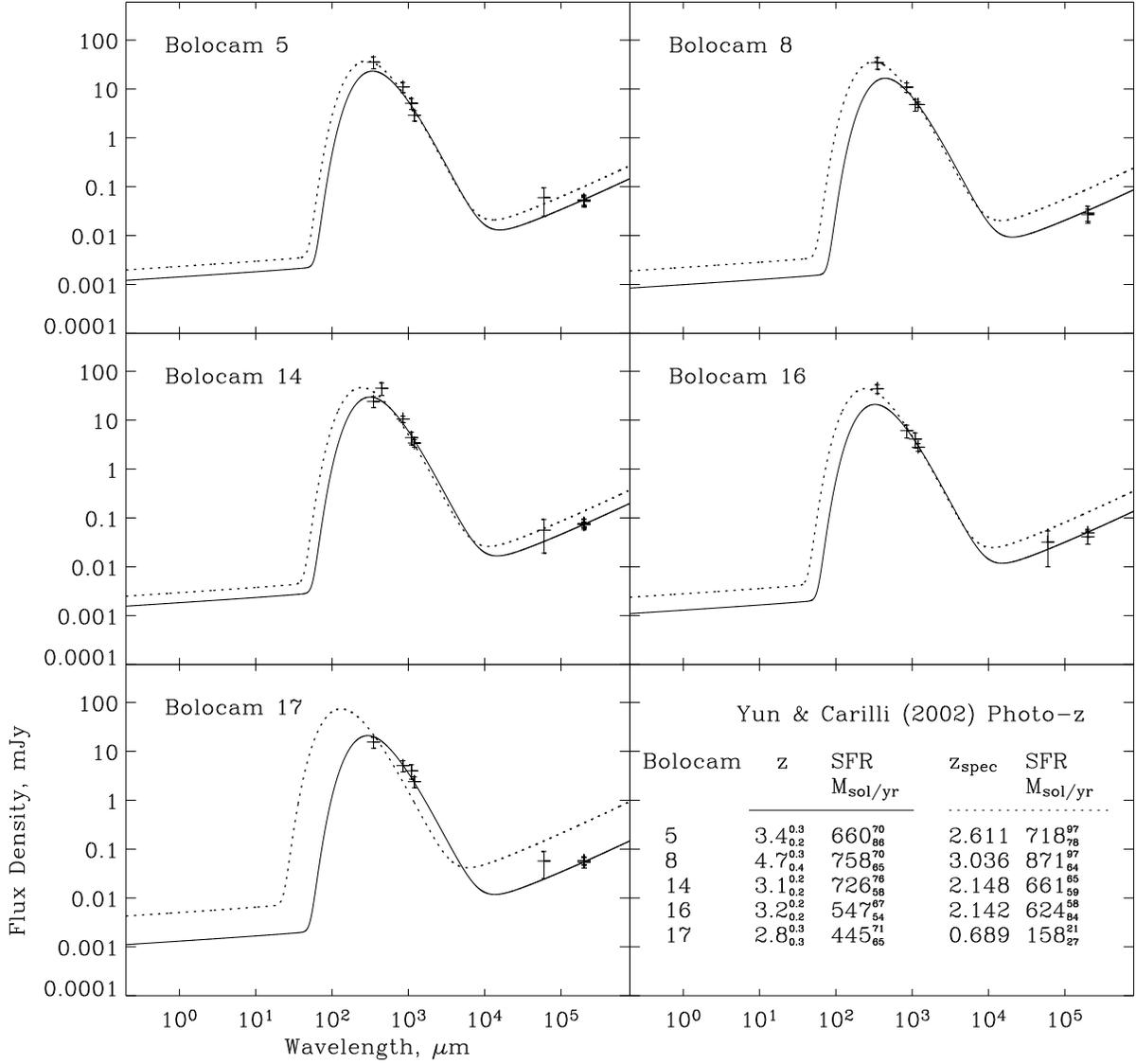}
\caption{Photometric redshift fits to five of the Bolocam galaxy candidates, using the model of \cite{yun02}.  The fits were limited to the submillimeter, millimeter and radio 
points shown.  The solid line represents a two parameter fit, allowing both $z$ and SFR to vary.  The dashed line is the result of fixing the redshifts to the 
spectroscopic redshifts of \cite{chapman05}, normalized to the submillimeter points.  The best fit redshifts and SFRs, along with their error bars (from Monte Carlo simulations) are listed for each 
source.}
\label{figure:photoz}
\end{figure}

2) Entire FIR-Radio SED -- \cite{yun02} utilized the entire Far-IR to radio spectral energy distribution to estimate photometric redshifts and SFRs.  The 
redshift template is based upon the theoretical models of thermal dust emission, thermal bremsstrahlung (free-free) emission, and 
nonthermal synchrotron emission for dusty starburst galaxies.  Photometric redshift fits of the five Bolocam galaxy candidates (5, 8, 14, 
16, 17) with spectroscopic redshifts \citep{chapman05} are shown in Figure \ref{figure:photoz}, 
with best-fit redshifts (and errors) also listed in Table \ref{table:redshifts_yunandcarilli}.  The solid lines in the figure represent 
the best fit spectrum to the submillimeter, millimeter, and radio point shown.  The dotted line 
represents a second fit using the \cite{yun02} model, this time fixing the spectroscopic redshift and normalizing (varying only the SFR) to the submillimeter 
points.  

\begin{deluxetable}{ccccccccccc}
\tabletypesize{\tiny}
\tablecaption{Photometric Redshifts}
\tablewidth{0pt}
\tablehead{
\colhead{Bolocam} & \colhead{N$_{\mathrm{Radio}}$} & \colhead{N$_{\mathrm{Submm}}$} & \colhead{N$_{\mathrm{{\it Spitzer}}}$} & 
	\colhead{Carilli \& Yun} & \colhead{Yun \& Carilli} & \colhead{Wiklind} 	& \colhead{Laurent} 		& \colhead{MRK231} & \colhead{ARP220} & \colhead{Chapman}\\
\colhead{1.1 mm} & \colhead{} & \colhead{} & \colhead{} & 
	\colhead{1999} & \colhead{2002} 		& \colhead{2003} 	& \colhead{et al.\ 2005} 	& \colhead{} & \colhead{} & \colhead{et al.\ 2005}\\
\colhead{Number} & \colhead{} & \colhead{} & \colhead{} & 
	\colhead{z$_{\mathrm{phot}}$} & \colhead{z$_{\mathrm{phot}}$} & \colhead{z$_{\mathrm{phot}}$} & \colhead{z$_{\mathrm{phot}}$} & \colhead{z$_{\mathrm{phot}}$} & \colhead{z$_{\mathrm{phot}}$} & \colhead{z$_{\mathrm{spec}}$}
}
\startdata
1	&2(S)	&3	&3	&4.6$^{+0.3}_{-0.3}$				&4.1$^{+0.3}_{-0.3}$				&3.2$^{+0.4}_{-0.5}$	&3.2$^{+0.3}_{-0.4}$	&15$^{+1.4}_{-1.5}$	&3.0$^{+0.2}_{-0.3}$	&\\
"	&1(N) 	&"	&3	&4.6$^{+0.3}_{-0.3}$				&4.1$^{+0.3}_{-0.3}$				&"			&"			&1.8$^{+1.0}_{-0.8}$	&2.0$^{+0.7}_{-0.7}$	&\\
2	&2,1(N,S)&2	&0	&2.2$^{+0.2}_{-0.2}$, 3.9$^{+0.3}_{-0.3}$	&0.6$^{+0.1}_{-0.1}$, 2.9$^{+0.7}_{-0.8}$	&4.2$^{+0.6}_{-0.8}$	&3.9$^{+0.4}_{-0.4}$	&			&			&\\
3	&2(NE)	&2	&5	&2.6$^{+0.2}_{-0.2}$				&0.7$^{+0.1}_{-0.1}$				&4.7$^{+0.8}_{-0.7}$	&4.2$^{+0.4}_{-0.5}$	&0.2$^{+0.2}_{-0.2}$	&3.1$^{+0.2}_{-0.2}$	&\\
"	&2(SW)	&2	&3	&3.7$^{+0.3}_{-0.3}$				&1.6$^{+0.3}_{-0.4}$				&4.9$^{+0.8}_{-0.7}$	&4.3$^{+0.5}_{-0.4}$	&10$^{+0.9}_{-0.8}$	&2.3$^{+0.7}_{-0.2}$	&\\
4	&1	&2	&0	&5.7$^{+0.7}_{-0.8}$				&5.1$^{+0.7}_{-0.9}$				&			&			&			&			&\\
5	&3(E)	&4	&5	&4.3$^{+0.4}_{-0.4}$				&3.4$^{+0.2}_{-0.3}$				&2.5$^{+0.5}_{-0.4}$	&2.6$^{+0.4}_{-0.5}$	&1.2$^{+0.2}_{-0.2}$	&2.6$^{+0.2}_{-0.2}$	&2.611\\
6	&2,2(E,W)&2(E)	&0	&4.2$^{+0.5}_{-0.4}$, 3.1$^{+0.3}_{-0.3}$	&3.5$^{+0.4}_{-0.5}$, 2.0$^{+0.3}_{-0.3}$	&3.2$^{+0.7}_{-0.7}$	&3.2$^{+0.5}_{-0.6}$	&			&			&\\
"	&2	&2(W)	&0	&1.6$^{+0.2}_{-0.2}$				&0.8$^{+0.1}_{-0.1}$				&1.9$^{+0.4}_{-0.5}$	&2.0$^{+0.5}_{-0.6}$	&			&			&\\
7	&1	&1	&0	&4.2$^{+0.6}_{-0.8}$				&3.6$^{+0.6}_{-1.0}$				&			&			&			&			&\\
8	&2(N)	&4	&4	&5.1$^{+0.5}_{-0.6}$				&4.7$^{+0.3}_{-0.3}$				&3.1$^{+0.4}_{-0.6}$	&3.1$^{+0.3}_{-0.5}$	&0.6$^{+0.3}_{-0.3}$	&2.4$^{+0.6}_{-0.2}$	&\\
"	&1(S)	&"	&4	&5.1$^{+0.6}_{-0.6}$				&4.7$^{+0.4}_{-0.4}$				&"			&"			&0.7$^{+0.4}_{-0.1}$	&3.1$^{+0.2}_{-0.6}$	&3.036\\
9	&2	&1	&5	&3.9$^{+0.4}_{-0.4}$				&3.2$^{+0.5}_{-0.5}$				&			&			&1.3$^{+0.2}_{-0.2}$	&1.8$^{+0.2}_{-0.5}$	&\\
10	&0	&1	&0	&						&						&			&			&			&			&\\
11	&1	&1	&2	&3.2$^{+0.4}_{-0.5}$				&2.5$^{+0.4}_{-0.6}$				&			&			&0.4$^{+0.2}_{-0.2}$	&0.8$^{+0.3}_{-0.2}$	&\\
12	&0	&1	&0	&						&						&			&			&			&			&\\
13	&2,1(NE,SE)&1	&0	&3.9$^{+0.4}_{-0.4}$, 4.8$^{+0.6}_{-0.6}$	&3.2$^{+0.5}_{-0.4}$, 4.4$^{+0.8}_{-0.8}$	&			&			&			&			&\\
" 	&1(E)	&"	&4	&5.0$^{+0.6}_{-0.7}$				&4.6$^{+0.8}_{-1.0}$				&			&			&0.7$^{+0.3}_{-0.2}$	&2.4$^{+0.3}_{-0.5}$	&\\
14	&4	&5	&5	&3.7$^{+0.4}_{-0.3}$				&3.1$^{+0.2}_{-0.2}$				&3.2$^{+0.3}_{-0.4}$	&3.2$^{+0.3}_{-0.3}$	&7.3$^{+0.6}_{-0.2}$	&3.0$^{+0.2}_{-0.3}$	&2.148\\
15	&1	&1	&5	&4.1$^{+0.5}_{-0.4}$				&3.5$^{+0.6}_{-0.5}$				&			&			&0.6$^{+0.2}_{-0.2}$	&1.0$^{+0.2}_{-0.2}$	&\\
16	&2 	&4	&5	&4.1$^{+0.5}_{-0.5}$				&3.2$^{+0.2}_{-0.2}$				&1.9$^{+0.3}_{-0.5}$	&2.0$^{+0.4}_{-0.4}$	&0.4$^{+0.2}_{-0.2}$	&2.3$^{+0.3}_{-0.2}$	&2.142\\
17	&3(SE)	&4	&10	&3.8$^{+0.5}_{-0.4}$				&2.8$^{+0.3}_{-0.3}$				&3.4$^{+0.5}_{-0.7}$	&3.3$^{+0.4}_{-0.4}$	&3.4$^{+0.3}_{-0.3}$	&3.1$^{+0.2}_{-0.2}$	&0.689\\
" 	&1(NW)	&"	&5 	&5.3$^{+0.7}_{-0.8}$				&4.1$^{+0.4}_{-0.5}$				&"			&"			&0.3$^{+0.2}_{-0.2}$    &3.0$^{+0.2}_{-0.6}$	&\\
\enddata
\tablecomments{Best fit photometric redshifts of the Bolocam galaxy candidates using the models of \cite{carilli99}, \cite{yun02}, \cite{wiklind03}, and \cite{laurent05}, and 
cool and warm ULIRGs Arp 220 and MRK 231.  
N$_{\mathrm{Radio}}$, N$_{\mathrm{Submm}}$, and N$_{\mathrm{{\it Spitzer}}}$ are the number of coincident radio, submillimeter and {\it Spitzer} infrared points, respectively.}
\label{table:redshifts_yunandcarilli}
\end{deluxetable}

\begin{figure}
\epsscale{1.0}
\plotone{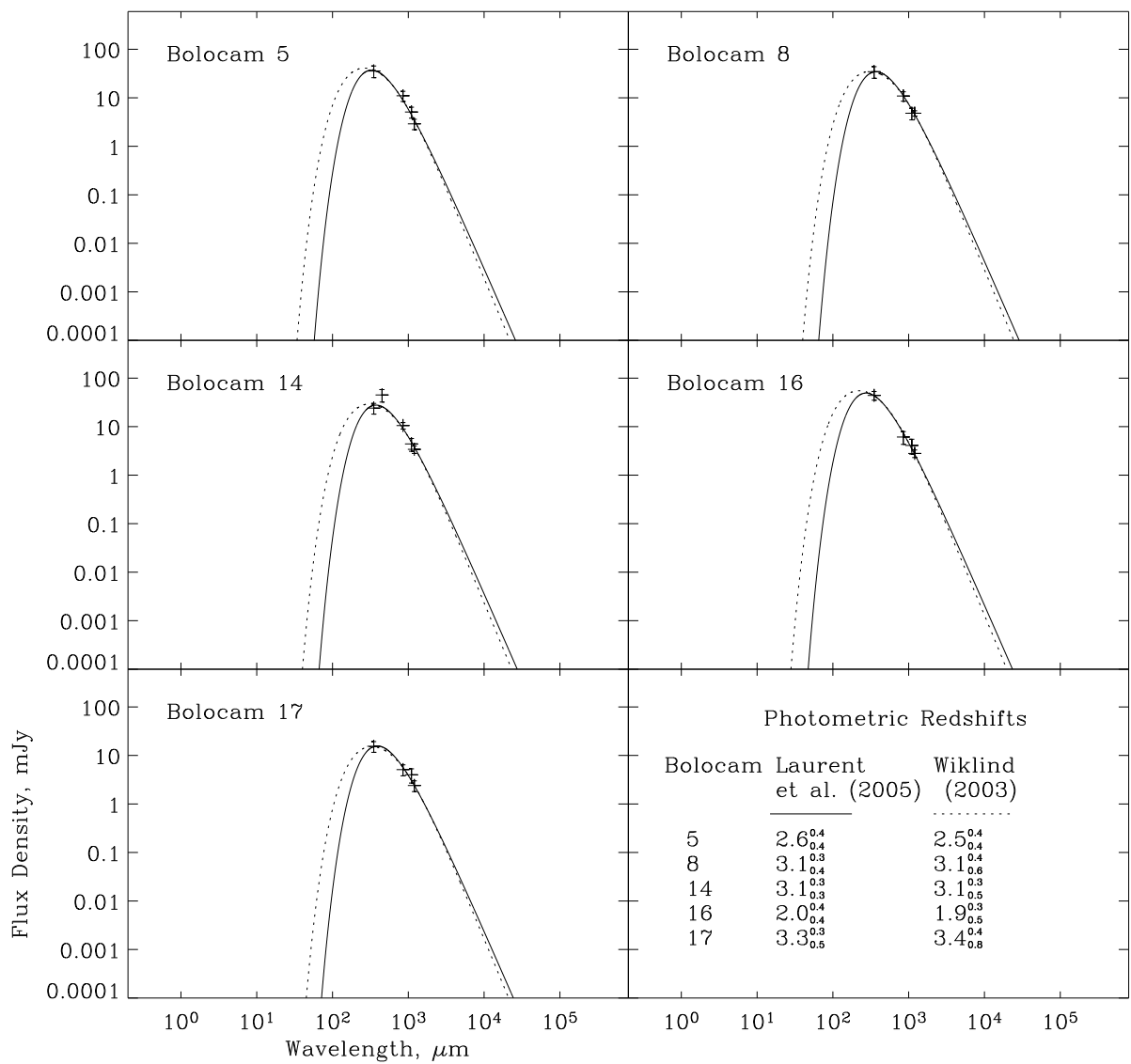}
\caption{Photometric redshift fits to five of the Bolocam galaxy candidates, using the model SEDs of Laurent et al.\ (2005, solid line) and Wiklind (2003, dotted line).  
The fits were limited to only the submillimeter and millimeter points shown.  The best fit redshifts, along with their error bars (from Monte Carlo simulations) are listed for each source.}
\label{figure:wiklind}
\end{figure}

3) Modified Blackbody -- \cite{wiklind03} found that observations of local ULIRGs exhibit a remarkably low dispersion in the far-IR to millimeter wavelengths 
($\lambda > 50 \mu$m), independent of whether the power source of the thermal emission is due to AGN or intense star formation.  
\cite{wiklind03} fit a simple blackbody emission spectrum (modified by a dust emissivity term) to sample of 37 local ULIRGs from \cite{klaas01}:
\begin{eqnarray}
\label{equation:greybody}
f_\nu \propto \epsilon_\nu B_\nu(T) \propto [1-\exp{(-\tau_\nu)}] B_\nu(T),
\end{eqnarray}
where $B_\nu(T)$ is the Planck function evaluated at dust temperature, $T$, and frequency, $\nu$, and $\tau_\nu$ is the optical depth 
of the dust:
\begin{eqnarray}
\nonumber
\tau_\nu = \left(\frac{\nu}{\nu_0}\right)^\beta.
\end{eqnarray}
\cite{wiklind03} made no assumption about the Wien side of the spectrum, as only the submillimeter ($\ge$ 450 $\mu$m) and millimeter points were fit.  Using the best-fit parameters from 
\cite{wiklind03}: $\beta = 1.8$, $\nu_0 = 1.2 \times 10^{12}$ Hz (250 $\mu$m), and $T_d = 
68$ K, we fit photometric redshifts to the SHARC II 350 $\mu$m, SCUBA 450 and 850 $\mu$m, Bolocam 1.1 mm, and MAMBO 1.2 mm flux densities 
of the galaxies detected in our Bolocam survey.  The two parameter fit (redshift and overall flux density normalization) yields redshifts for 9 of the 17 bolocam 
galaxies with $\ge$ 2 submillimeter/millimeter points.  Seven of the Bolocam galaxies (Bolocam 7, 9, 10, 11, 12, 13, 15) have no 
counterpart in the submillimeter/millimeter and one (Bolocam 4) has detections only at 1.1 and 1.2 mm, which is an insufficient 
wavelength spread in order to properly constrain the galaxy redshift using this two parameter model.  
Redshift results for each of the 9 Bolocam galaxies are listed in Table \ref{table:redshifts_yunandcarilli}.  The best fit models to 
the five galaxies with \cite{chapman05} redshifts (Bolocam 5, 8, 14, 16, 17) are shown in Figure \ref{figure:wiklind}.  

Similar to the method of \cite{wiklind03}, \cite{laurent05} created a composite SED of nearby dusty {\it IRAS} galaxies, 
high-redshift submillimeter galaxies, gravitationally lensed high-redshift
galaxies, and high-redshift AGNs \citep[][and references therein]{blain02}, and found fit parameters of $T$ = 40 K, $\nu_0$ = 3700 
GHz, and $\beta$ = 1.6 for Equation \ref{equation:greybody}.  Redshift results for each of the 9 Bolocam galaxies are listed in Table 
\ref{table:redshifts_yunandcarilli}.  The best fit models to the five galaxies with \cite{chapman05} redshifts (Bolocam 5, 8, 14, 
16, 17) are shown in Figure \ref{figure:wiklind}.

\begin{figure}
\epsscale{1.0}
\plotone{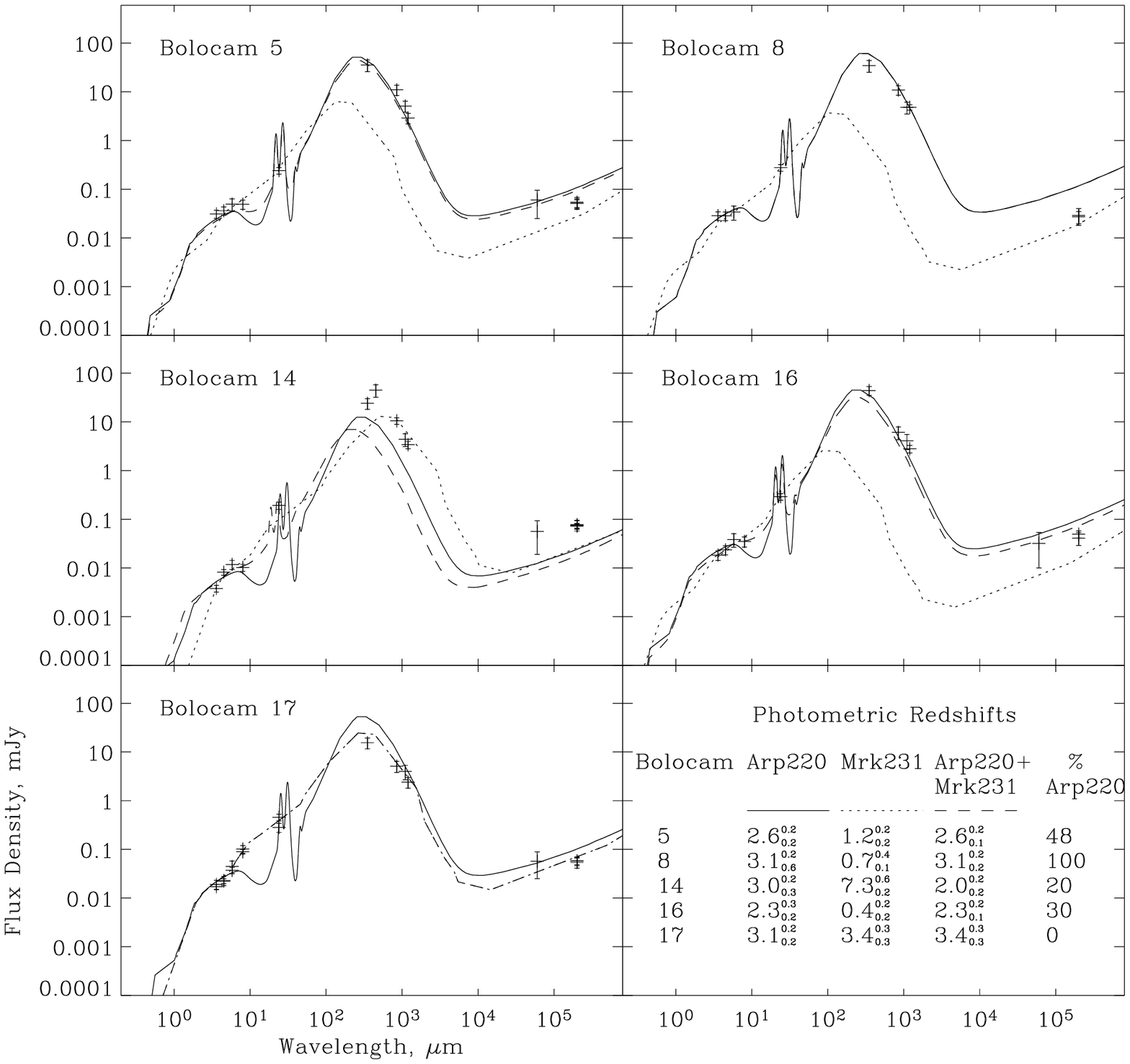}
\caption{Photometric redshift fits to five of the Bolocam galaxy candidates, fitting the {\it Spitzer} near- and mid-infrared points to \cite{egami04} models of a cool ULIRG
Arp 220 \citep[solid line, from][]{silva98} and a warm (dominated by an AGN) ULIRG Mrk 231 (dotted line).  The (sub)millimeter and radio points were not used in the fit and are shown only 
for reference.  The dashed line represents a linear combination of the two models).  The best fit redshifts along with their error bars (from Monte Carlo simulations) are listed for each source.}
\label{figure:spitzer}
\end{figure}

4) Near IR Stellar Bump -- \cite{egami04} used the multiband imaging capabilities of the IRAC and MIPS IR cameras of the {\it Spitzer Space Telescope} to 
observe 38 VLA radio sources in the Lockman Hole.  They classified the resulting IR SEDs into two types: those showing a clear near-IR 
stellar continuum hump at a rest wavelength of 1.6 $\mu$m (due to the minimum opacity of the H$^-$ ion at 1.6 $\mu$m from photo-detachment and free-free transitions, which results in 
a local maximum in the the SEDs of cool stars, Sawicki 2002), and those with a featureless power-law continuum (from AGN).  We fit the {\it 
Spitzer} IR counterparts of the Bolocam galaxies with each of these spectra, using the \cite{egami04} models of a cool ULIRG 
Arp 220 \citep[from][]{silva98} and a warm (dominated by an AGN) ULIRG Mrk 231.  Only two fit parameters were used: the redshift, and 
an overall normalization.  Photometric redshifts were fit for each of the 12 Bolocam galaxies (Bolocam 1, 3, 5, 8, 9, 10, 
11, 13, 14, 15, 16, 17) with $\ge$ 2 IR {\it Spitzer} points.  Three Bolocam galaxies (Bolocam 2, 6, 12) were outside of the field 
surveyed by \cite{egami04}.  Two galaxies (Bolocam 4, 7) have {\it Spitzer} counterparts, but due to the high density of {\it 
Spitzer} sources in the field, they could not be uniquely associated with the Bolocam sources (because of a lack of another coincident 
detection in the submillimeter and/or radio).  The best-fit photometric redshifts for both Arp 220 and Mrk 231 (fitting only the {\it Spitzer} near- and mid-infrared points) are shown in 
Figure \ref{figure:spitzer}, with the resulting redshifts also listed in Table \ref{table:redshifts_yunandcarilli}. 

\subsection{Comparison of Photometric Redshift Techniques}
\label{subsection:discussion}

\begin{figure}
\epsscale{0.6}
\plotone{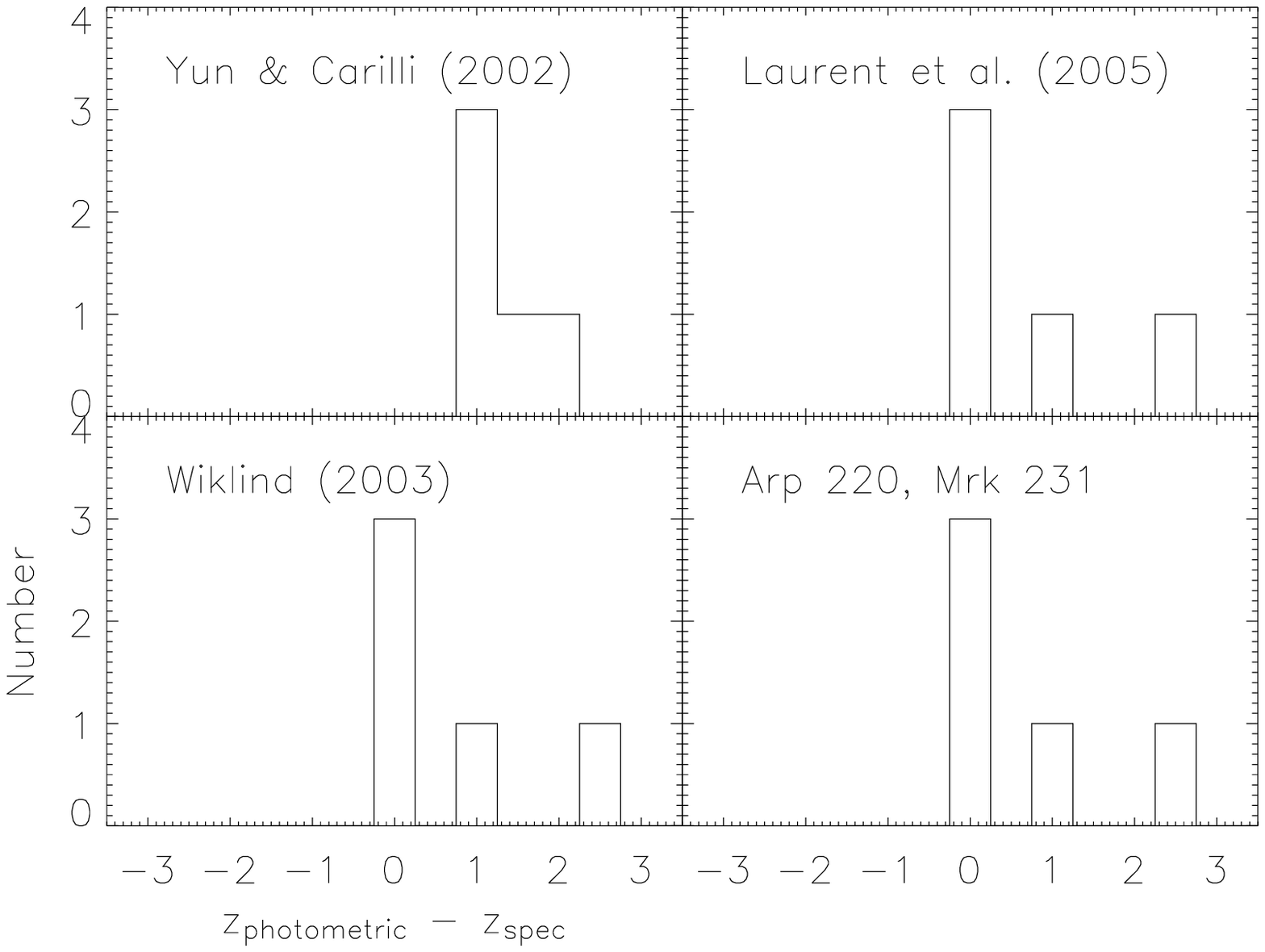}
\caption{Histogram of photometric redshift errors for various models compared to the spectroscopic redshifts of \cite{chapman05}.  \cite{yun02} assume 
the far-IR to radio correlation while both \cite{wiklind03} and \cite{laurent05} model the submillimeter and millimeter portions of the spectrum.  The lower right frame is a fit of the 
near- and mid-infrared points to models of Arp 220 and Mrk 231 \citep{egami04}.}
\label{figure:histogram}
\end{figure}

Comparing the results of each of the photometric redshift techniques with the spectroscopic redshifts of \cite{chapman05}
yields widely varying degrees of agreement.  Histograms of redshift errors for each of the photometric redshift models \citep[other than][in which coincident radio detections were 
treated separately]{carilli99} are shown in Figure \ref{figure:histogram}.  The histogram from fitting models of Arp 220 and Mrk 231 to the {\it Spitzer} IRAC and MIPS 
observations are the fits that result in the lowest residual $\chi^2$ (Arp 220 for Bolocam 5, 8, 14, and 16, and Mrk 231 for Bolocam 17).

The \cite{yun02} model \citep[as well as][]{carilli99} yields systematically high photometric redshifts compared to the spectroscopic redshifts by \cite{chapman05}.  
The comparison of model SEDs for photometric (solid line) and spectroscopic (dotted line) redshifts in Figure \ref{figure:photoz} suggests that SED
data points at the extreme ranges of wavelength coverage strongly influence the model fit and that the systematic tendency to derive a high redshift is primarily
driven by the lower than expected radio continuum flux density.  This is supported by the fact that fits to only the submillimeter and millimeter-wave points yield much more accurate photometric 
redshifts (see below).  This is perhaps not surprising, given recent evidence \citep{chapman05} suggesting a large degree of dispersion in the radio-to-far-IR correlation at higher redshift. 
Nevertheless, it is unlikely that the \cite{yun02} model template, which is derived from the ensemble average of 23 infrared luminous galaxies in the local universe, is systematically biased by radio 
bright objects because $\ge$98\% of all FIR-selected galaxies follow the well known and tight radio-FIR correlation, independent of FIR luminosity \citep{yun01}.  Aside from Bolocam source 17 whose 
spectroscopic redshift by  \cite{chapman05} appears suspect (see \S\ \ref{subsection:bolocam17}), these comparisons suggest that the observed radio continuum in Bolocam galaxies is 2-5 times 
fainter than predicted by the synchrotron flux densities (which dominate thermal brehmsstrahlung by a factor of $\sim$ 13 at 20 cm) from the low-redshift ULIRGs 
for which the local FIR-radio correlation was derived (see further discussions in \ref{subsection:radio}).  


In contrast to the far-IR-to-radio correlation photometric redshift techniques, both the \cite{wiklind03} and \cite{laurent05} 
modified blackbody curves correctly estimate the redshifts for three of the five Bolocam galaxies with spectroscopic redshifts (within the uncertainties of the photometric redshift techniques).  The 
strength of the submillimeter / millimeter-only photometric redshift technique is twofold.  First, while abandoning the radio 
points limits the number of points (as few as two, in some cases) to which we may fit a model, we ensure that the physics 
that dominates the region of the spectrum to which we are fitting is directly relevant to star formation-heated dust emission. 
Synchrotron radiation, by contrast, is dependent upon high energy electrons streaming through interstellar magnetic fields, whose
properties may vary as a function of environment (e.g.\ inverse-Compton losses for cosmic rays with higher CMB energy densities at high redshift -- see \S\ \ref{subsection:radio}).  Indeed, these 
galaxies are typically at least an order of magnitude more luminous than the low-redshift, infrared-luminous
galaxies from which the FIR-radio correlation was derived.

\begin{figure}
\epsscale{1.0}
\plotone{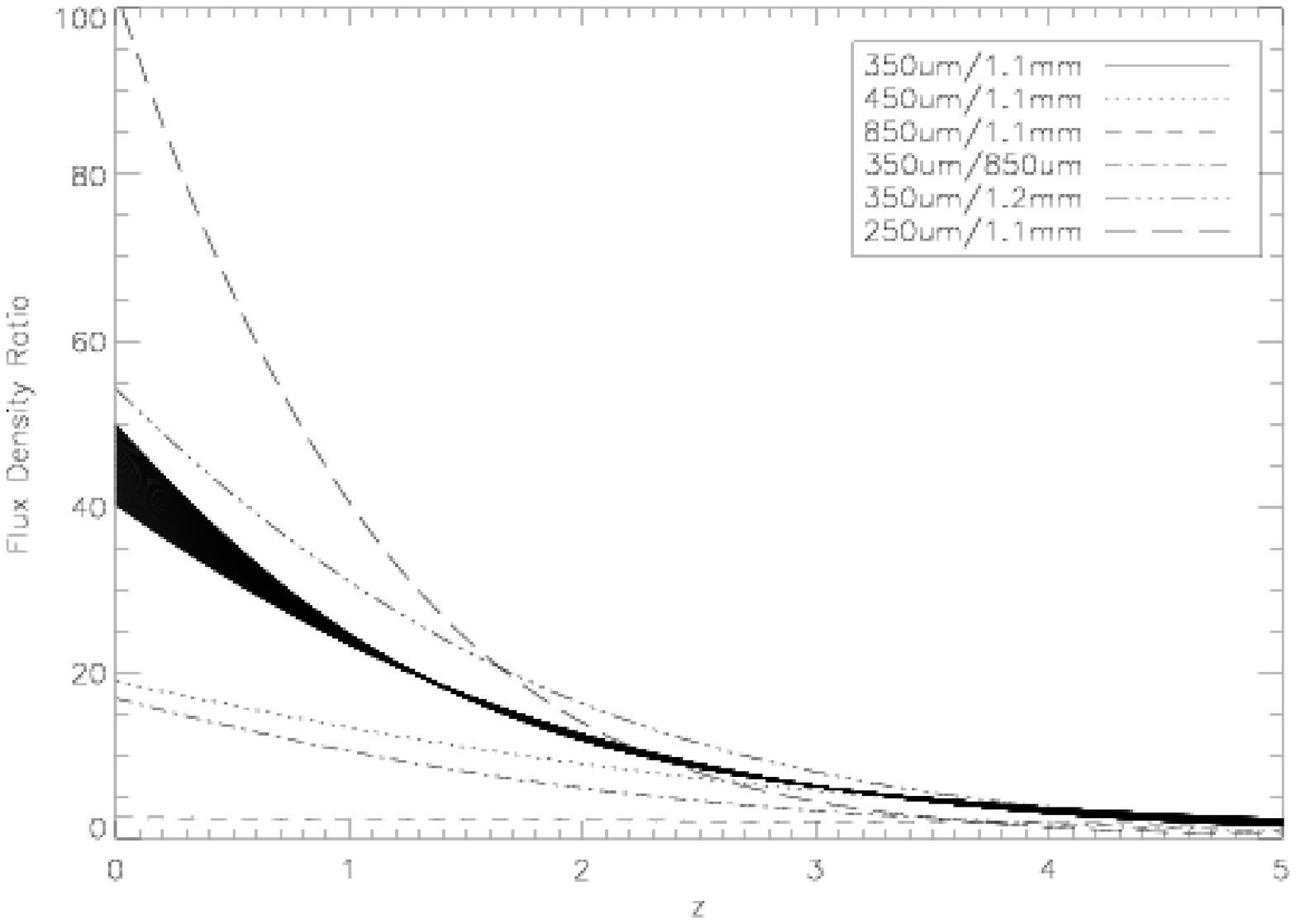}
\caption{Flux density ratio between various wavebands assuming the submillimeter galaxy model SED of \cite{laurent05}.  The model is based on the composite SED of nearby dusty IRAS 
galaxies, high-redshift submillimeter galaxies, gravitationally lensed high-redshift galaxies, and high-redshift AGNs \citep[][and references therein]{blain02}.  
The \cite{laurent05}
model assumes dust parameters of $T$ = 40 K, $\nu_0$ = 3700 GHz, $\beta$ = 1.6, and a power law of $f_\nu \propto \nu^{-1.7}$ to model the hotter dust components on the Wien side of 
the spectrum.  The shaded line represents the ratio of the SHARC 350 $\mu$m and Bolocam 1.1 mm wavebands, with the thickness corresponding to the difference between the \cite{laurent05}
and \cite{wiklind03} model SEDs.}
\label{figure:fluxratio}
\end{figure}

Second, having the Bolocam 
1.1 mm flux densities on the Rayleigh-Jeans side of the spectrum and the SHARC II 350 $\mu$m flux densities near the peak of the SED makes 
the 350 $\mu$m/1.1 mm flux density ratio a strong function of redshift.  This can be seen in Figure \ref{figure:fluxratio}, which 
shows the flux density ratios between various wavebands based on the \cite{laurent05} model SED.  The importance of the 
SHARC II 350 $\mu$m waveband is apparent.  For intermediate to high ($z < 5$) redshifts, the SHARC II flux density drops rapidly as 
a power law ($\sim \nu^{-1.7}$ due to the hotter components of dust) with redshift on the Wien side of the spectrum, while the millimeter-wave climbs up the steep 
Rayleigh-Jeans portion of the SED.  Photometric redshifts using the 450 or 850 $\mu$m wavebands are less sensitive than the 
350 $\mu$m / 1.1 mm wavebands.  Extending this analysis shows the discriminatory power of the BLAST and {\it 
Herschel} space telescope 250 $\mu$m bands in conjunction with a millimeter waveband, although the far-IR waveband begins to probe a range of hotter dust temperatures.  (The flux density ratio of 
these wavebands may not be well correlated, as discussed in \S\ \ref{subsection:irspectrum}.)  BLAST \citep[Balloon-borne Large-Aperture Submillimeter Telescope,][]{devlin01} is a 
balloon-based instrument which incorporates a 2-meter primary mirror and is equipped with large-format bolometer cameras operating at 250, 350, and 500 $\mu$m which, when complete, will 
provide the first sensitive large-area (0.5-40 deg$^2$) submillimeter surveys at these wavelengths.  The bolometer arrays are prototypes of the Spectral and Photometric Imaging Receiver 
(SPIRE) focal plane cameras for the {\it Herschel} satellite \citep{griffin01}, which will further investigate the formation and evolution of AGNs and star formation in 
high redshift submillimeter galaxies. 

It is important to note that while both the \cite{wiklind03} and \cite{laurent05} modified blackbody curves both produce reasonably accurate photometric redshifts (and produce nearly identical SEDs on 
the Rayleigh-Jeans portion of the spectrum), they make substantially different assumptions about the dust properties:  $T_{\mathrm{dust}}$ and $\beta$ are 68 K and 1.8 for the \cite{wiklind03} 
model SED and 40 K and 1.6 for the \cite{laurent05} model SED.  This points to the degeneracy of the dust temperature and the grain emissivity index.  While the shapes of the submillimeter SEDs are 
reasonably modeled by either dust model and thus predict photometric redshifts with some accuracy, essentially no information about the dust temperatures can be inferred.  In fact, 
representing the dust SED with two (or more) components produces similar $\chi^2$ values (and thus similar redshifts), with both temperatures {\it lower} than that of the single dust 
temperature model \citep{wiklind03}.  

The photometric redshifts determined by fitting the {\it Spitzer} infrared observations yield redshifts that are equivalent to both the \cite{laurent05}
and \cite{wiklind03} model SEDs.  This confirms the conclusions of \cite{egami04}, in which starburst-dominated galaxies ("cold") show remarkably similar SEDs 
in the infrared.  The resulting photometric redshifts are highly sensitive to the 1.6 $\mu$m continuum hump (and PAH features), with a sharp minimum in $\chi^2$.  
(The ARP 220 fits may be biased towards particularly good fits as 4 of the 5 galaxies with spectroscopic redshifts lie between 2 $\lesssim z \lesssim$ 3, which is optimal 
for the 8 $\mu$m PAH feature to be shifted into the observed 20 $\mu$m IRAC waveband.)  AGN-dominated ("warm") galaxies also show very similar SEDs, but lacking a strong 
continuum feature in the infrared, are subject to larger redshift fitting uncertainties; a brighter, higher redshift galaxy is 
characterized by a similar shape in the infrared portion of the SED as a cooler, low-redshift galaxy, with the Wien side of the spectrum well-modeled with a power-law \citep{blain99}.  


\subsection{Bolocam Source 17: Spectroscopic Misidentification?}
\label{subsection:bolocam17}

We bring special attention to photometric redshift analysis of Bolocam source 17 (corresponding to SCUBA source 8), as the \cite{laurent05}, \cite{wiklind03}, and {\it Spitzer} IR 
models are consistent in overpredicting the redshift of this galaxy ($z_{\mathrm{spec}}$ = 0.689) by $\ge 4 \sigma$ ($z_{\mathrm{phot}}$ = 3.3$^{0.4}_{0.5}$, 3.4$^{0.6}_{0.6}$, and 
3.4$^{0.3}_{0.3}$, respectively).  We point out that the large offset between the spectroscopic and photometric redshifts is possibly the result of source confusion, as 
two radio sources (both with {\it Spitzer} and optical counterparts) fall near the center of the Bolocam, SCUBA, MAMBO, and SHARC II error circles, within 4$\arcsec$ of each other.  (Egami 
et al.\ 2004 
refer to the northwest and southeast radio sources as LE850.8a and LE850.8b, respectively.)  The northwest 
radio source is believed by \cite{lehmann01} to be the counterpart to the ROSAT X-ray emission, who find a redshift of 0.974 using optical Keck spectroscopy.  Using XMM-Newton 
observations, however, \cite{ivison02} conclude that the X-ray source corresponds to the southeast radio source.  Indeed, the linear fit of a combination of ARP 220 and MRK 231 ULIRG models to the 
{\it Spitzer} points coincident with the southeast radio source yields a 100\% warm (AGN dominated) component.  It is near this radio position that \cite{chapman05} find a 
spectroscopic redshift of 0.689.  In fact, the X-ray emission observed with both instruments appears to fall between these two radio sources.  Optical R-band images from both \cite{yun05}, and 
\cite{ivison05} show multiple optical counterparts at the southeast radio source position.  Furthermore, the spectroscopic position quoted by \cite{chapman05}
appears to coincide with an optical source $\sim$ 2$\arcsec$ to the south of the southeast radio source, a source detected with four-color SDSS photometry (in addition to Yun et 
al.\ 2005 and Ivison et al.\ 2005 R-band photometry) and cataloged as a low-redshift galaxy.  We conclude that it is possible that the submillimeter detections may either be suffering from 
source confusion from two or more galaxies, or that the \cite{chapman05} redshift corresponds to a source other than that of the southeast radio detection.  If the latter is true, 
then the consistent redshifts predicted by the Laurent et al./Wiklind (2005, 2003) and {\it Spitzer} IR models may further point to the accuracy of these photometric 
redshift techniques.

\section{Discussion}

\subsection{IR Spectrum}
\label{subsection:irspectrum}

\begin{figure}
\epsscale{1.0}     
\plotone{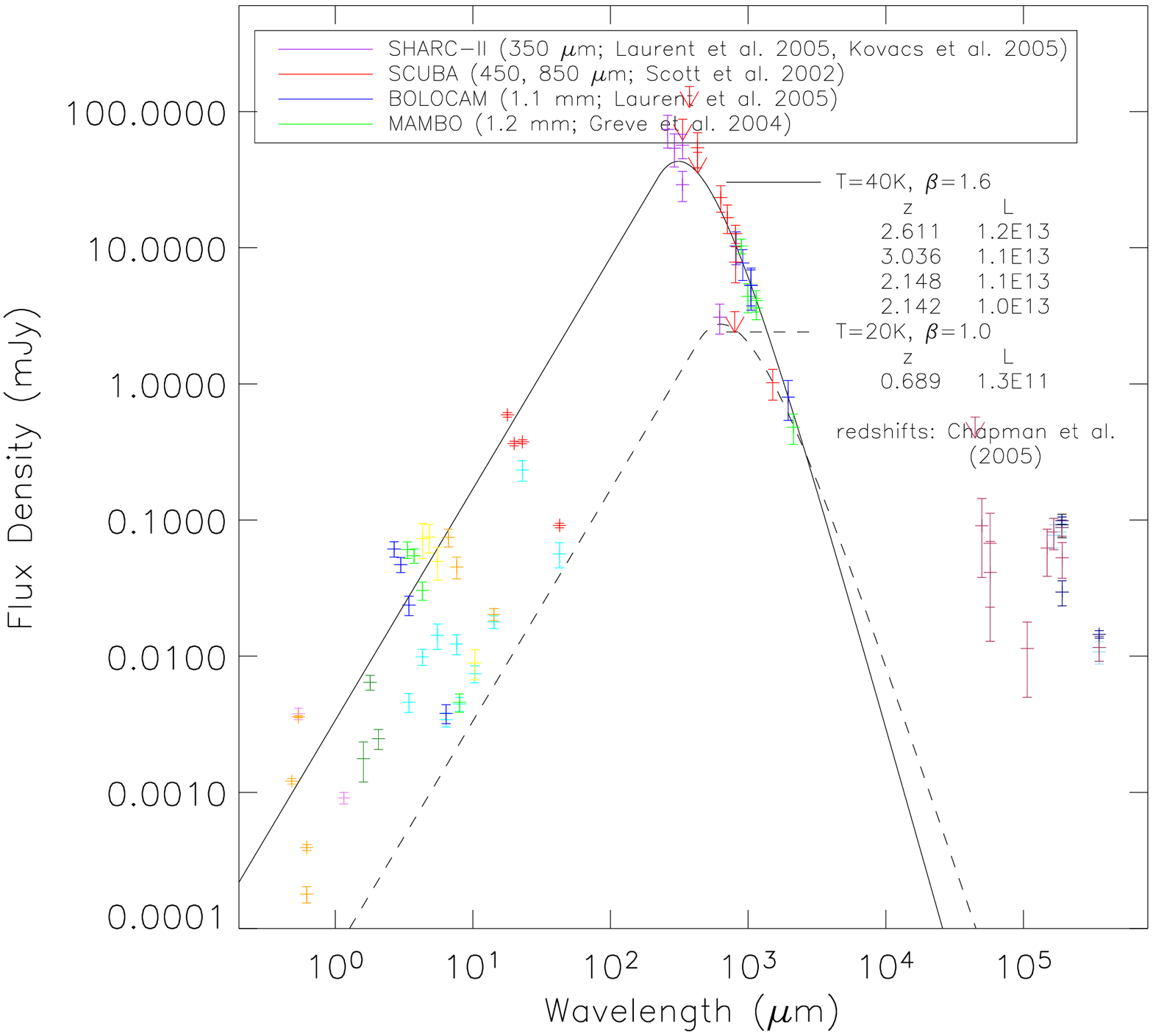}
\caption{Composite SED for the five Bolocam galaxies with spectroscopic redshifts (Bolocam 5, 8, 14, 16, and 17).  The observed SEDs have all been redshifted to $z$ = 2.0 using the 
spectroscopic redshifts, corrected for cosmological dimming, and their Bolocam 1.1 mm fluxes normalized to the \cite{laurent05} model.  The solid line represents the \cite{laurent05} model (T = 40 K, 
$\beta$ = 1.6) based on the composite SED of Blain et al.\ (2002, and references therein).  The dashed line represents the same model, with T = 20 K and $\beta$ = 1.0.} 
\label{figure:revert}
\end{figure}

SEDs over the entire IR-radio spectral range of the five Bolocam galaxies with spectroscopic redshifts are shown in Figure 
\ref{figure:revert}.  These spectra have the same redshift, cosmological dimming, and normalization corrections as in Figure 
\ref{figure:revertzoom}.  While four of the five galaxies have closely correlated spectra in the submillimeter region of the spectrum, 
the infrared spectra ({\it Spitzer} 3.6, 4.5, 5.8, and 8.0 $\mu$m IRAC and 24 $\mu$m MIPS observations) exhibit a large degree of 
dispersion.  This dispersion may be the 
result of several things: 1) Because the spectra have been normalized to a T = 40 K ($\beta=1.6$) spectrum at their 1.1 mm Bolocam flux densities to 
account for intrinsic brightness variation between the galaxies, the normalization will result in an artificial reduction in the 
submillimeter flux density dispersions.  This effect is not likely to dominate, as the flux density normalization has a $\sim$ 10\% effect on the flux densities 
of the galaxies.  2) The {\it Spitzer} detection associated with Bolocam source 14 lies systematically low compared to the other three 
galaxies well-modeled by a T=40 K dust spectrum.  While these {\it Spitzer} observations \citep{egami04} are from a different data 
set than the remaining {\it Spitzer} observations (this work), it is unlikely that their flux densities are 
systematically uncertain by nearly an order of magnitude.  3) The 40 K dust temperature model of the submillimeter portion of the 
spectrum for Bolocam sources 5, 8, 14, and 16 assume a single dust temperature for each of the four sources.  The temperature fit 
parameter is somewhat degenerate with other fit parameters, including the critical frequency, $\nu_0$, where the optical depth of the 
dust is unity.  Thus, if Bolocam source 14 has a lower characteristic dust temperature, then the infrared portion of the 40 K model will 
significantly overestimate the infrared flux density.  \cite{chapman05} estimates the temperature of Bolocam source 14 to be 33 K from two 
photometric points (850 $\mu$m SCUBA and 1.4 GHz VLA radio observations) and the dust SED templates of \cite{dale02}.  This 
temperature uncertainty likely dominates our uncertainty in matching the infrared flux densities.  4) In addition to heating by the 
ultraviolet and optical flux density from young stars associated with ongoing star formation, the thermal dust emission responsible 
for the bright submillimeter flux densities may be contributed to by an energetic AGN.  While not dominating the total bolometric output from 
the galaxy, they may have a non-negligible (20\%) contribution \citep{alexander05}.  If this is the case, then the shape of the 
infrared continuum may be vary according to the relative contribution of star formation rates for these galaxies.  5)  Another possible 
explanation for the larger dispersion in {\it Spitzer} infrared flux density as compared to our single dust temperature model may be from the fact 
that the infrared flux densities trace separate epochs of star formation within the galaxy.  It is plausible that the dust heated 
by the ultraviolet and optical flux density from from current star formation is not well correlated to the current infrared flux density of older stars 
(from previous star formation).  Furthermore, models of UV to millimeter emission of star clusters embedded in optically thick giant molecular clouds (GMCs) suggest that the near-infrared to 
far-infrared portion of starburst galaxy SEDs vary considerably with age of the starburst \citep{efstathiou00}.

\subsection{Radio Spectrum / FIR-Radio Correlation}
\label{subsection:radio}

The composite SED (including the radio points) of the 5 Bolocam sources with spectroscopic redshifts
is shown in Figure~\ref{figure:revert}.  Two interesting observations
can be made about the radio continuum emission associated with these
galaxies: (1) the radio continuum is lower (or the submillimeter continuum is higher) on average than the well
established radio-FIR correlation for the local universe by Yun, Reddy, \& Condon (2001, cf.\ \S\ \ref{subsection:discussion});
and (2) like the infrared flux densities, the 6 and 20 cm VLA radio flux densities 
for the four galaxies with closely correlated spectra in the submillimeter region of the 
spectrum show a large degree of dispersion (factor of 5 in 20 cm flux density).  This scatter is much larger than the 
quoted uncertainties of the VLA radio observations (which constrain the 20 cm radio flux density of each galaxy to better than 20\%) This degree of dispersion 
is also larger than the factor of 3 scatter in the radio-FIR correlation seen among the FIR selected galaxies in
the local universe.  In fact, while the spectroscopic redshifts are similar for the four galaxies (2.1 $< z <$ 3.0), 
varying the 20 cm flux density over the observed range causes the best fit photometric redshifts of \cite{yun02} 
to vary from $z = 2.7$ to 4.8.  This dispersion undoubtably contributes to the large errors of the photometric redshifts 
discussed in \S\ \ref{section:photometric_redshifts}.

Deep radio continuum imaging using the VLA is a
technically challenging task, and the disparate 20 cm flux densities 
by $\sim$ 50\% reported for Bolocam sources 2 and 6 (cf.\ \S\ \ref{section:summary}) 
exemplifies the difficulty of the photometry at radio wavelengths.
Most systematic noises in interferometry tend to suppress the brightness
of astronomical sources, and part of the lower radio continuum flux
density might be related to the imaging and photometry problems.

It is important to note that due to the low S/N ratios at which the sources have been detected 
in the submillimeter wavebands ($\le 4 \sigma$ for the vast majority of SHARC, SCUBA, Bolocam, and MAMBO detections), 
flux bias \citep{laurent05} plays a major role in overestimating the flux densities at these wavelengths.  
While this factor indeed results in a systematic shift of the entire submillimeter portion of the spectrum to higher flux densities, 
it is unlikely that the magnitude of this effect ($\sim$ 20\% for the Bolocam flux densities) could fully account for the 
lower than expected (from the FIR-radio correlation) radio continuum flux densities.  Calibration errors may also contribute to a systematic overestimate of the submillimeter flux densities. 
Finally, as two or more independent radio sources are found within the Bolocam error circle in 8 out of 17 cases in Figure 2, source confusion
or source blending may also contribute to the apparently lower radio continuum flux if only one radio source is identified as the counterpart.

The lower radio continuum flux density and the larger scatter may also
reflect an actual breakdown in the radio-FIR correlation. 
Inverse-Compton losses for the high energy cosmic rays responsible
for the synchrotron radiation is thought to be significant at $z > 2$,
and a possible breakdown in the radio-FIR correlation has been 
considered previously (see Condon 1992, Carilli \& Yun 2000).  
This effect is demonstrated in Figure \ref{figure:spitzer}, in which models
of local ULIRGs Arp 220 and Mrk 231 systematically overestimate the radio flux
densities of these submillimeter galaxies.  Higher quality data on a larger sample of high redshift systems are needed to
examine the importance of inverse-Compton loss and the possible breakdown in the radio-FIR correlation.

X-ray heating of the circum-nuclear gas and dust is an important source of
luminosity in the far-IR if a luminous AGN is present (Maloney, Hollenbach, \& Tielens 1996).
Radio-quiet AGN FIR emission could reduce the 1.4 GHz flux density with
respect to the FIR heating.  Alexander et al.\ (2005, and references
therein) make the case using X-ray detections and spectral indices that many, perhaps most,
submillimeter galaxies have AGN but that they are not bolometrically important.  However,
the statistics of X-ray detected submillimeter galaxies for which hard/soft ratios can be measured is not large and it cannot be ruled out that many submillimeter
galaxies are Compton thick (N$_\mathrm{H}$ $>$ 1.5 $\times$ 10$^{24}$ cm$^{-2}$).  
We conclude that the generally low radio flux densities in our sample could be due to small number statistics, source confusion, or
generally depressed radio emission, perhaps due to quenching of high energy cosmic rays,
although radio-quiet, Compton-thick AGN contributions to the dust heating cannot be ruled out.
There are new AGN versus star formation spectral diagnostics emerging (Ivison et al., Egami
et al.), and it is possible that to definitively settle the issue may ultimately require
ALMA, Constellation-X, and interferometric FIR spectral line diagnostic capability.

\subsection{Inferred Luminosities}

To obtain the intrinsic bolometric luminosities of the five Bolocam galaxies with spectroscopic redshifts, the spectra of 
\S\ \ref{subsection:correlations_between_spectra} (based on the Laurent et al.\ 2005 model) were 
integrated for each galaxy in their respective rest frames.  The resulting bolometric 
luminosities are listed in Figure 
\ref{figure:revertzoom}.  The four galaxies well-modeled by a 40 K dust spectrum have luminosities ranging from $L = (1.0-1.2)
\times 10^{13}$ $\mathrm{L}_\odot$.  The lower redshift galaxy (Bolocam source 17) has an inferred luminosity two orders of magnitude 
lower ($L = 1.3 \times 10^{11}$ $\mathrm{L}_\odot$).  If the spectroscopic redshift of 0.689 does not apply to this galaxy and it instead lies at $z$ = 3.4 (the photometric redshift predicted by the 
Laurent et al. 2005 / Wiklind 2003 and {\it Spitzer} IR models), then its luminosity of $L = 8.2 \times 10^{12}$ $\mathrm{L}_\odot$ agrees well with the others.

\subsection{Stellar and Dust Masses Implied from the Integrated Submillimeter Luminosities}

Three hundred and fifty micron observations combined with far-infrared and millimeter-wavelength photometry enables accurate measurements of luminosity for
galaxies near $z$ = 2 because no interpolation across the peak of the SED is required.  Characteristically, integration of the SEDs of
the galaxies in our sample from far-infrared to millimeter-wavelengths yields luminosities of $\sim$ 1 $\times$ 10$^{13}$ L$_\odot$.  Assuming: 1) that the 
luminosity derives from star formation (young stars, which may overestimate stellar masses due to contribution from intermediate mass giants), 
2) a characteristic (Salpeter) form of the initial 
mass function (IMF) from \cite{chabrier03}, and 3) all of the optical and ultraviolet radiation is reprocessed to long wavelengths by dust, enables the stellar mass content of the galaxies to be 
approximately estimated.  
%
We adopt the $M^{3.5}$ luminosity function of \cite{demircan91}.
Because the luminosity function is so steep, the derived mass depends strongly on the assumed upper mass limit of integration.   The lower limit of
integration is not well constrained by the data, although masses less than 0.3 M$_\odot$ are not likely to dominate the mass because the IMF flattens considerably at low
mass.  Lower mass limits of 0.7 and 1.0 M$_\odot$ could be relevant because:  1) \cite{dwek98} argued that a Salpeter IMF for $z$ $>$ 1 galaxies cannot
extend much lower than this without producing too many low mass stars that would be present today and 2) \cite{chabrier03} suggests (with caution) that the 
high-z IMF could cut off $>$ 1 M$_\odot$ based on multiple circumstantial lines of evidence.  Varying $m_l$ from 0.3 to 1.0 M$_\odot$ and limiting $m_u$ to $\le$ 50 M$_\odot$ yields a mininum 
stellar mass of 10$^{10}$ M$_\odot$ and a maximum of a few $\times$ 10$^{11}$ M$_\odot$, consistent with the stellar mass content of large elliptical galaxies, as previously pointed out by many
authors \citep[e.g.,][]{smail02, lilly99}.  This range must be considered an upper limit because AGN could be responsible for 
some of the dust heating.

For all of the galaxies with secure 350 $\mu$m detections, especially those with \cite{chapman05} spectroscopic redshifts, it is clear that the Bolocam 1.1 mm observations
lie on the optically thin Rayleigh-Jeans side of the SED, and therefore enable dust mass estimates.  The flux density, $S_\nu$, of a galaxy at an observed frequency, $\nu$, is related to the dust 
mass, M, by
\begin{eqnarray}
\nonumber
S_\nu=B_{\nu'}(T) \frac{\kappa_\nu M (1+z)}{D_L^2},
\end{eqnarray}  
where $D_L$ is the luminosity distance to redshift z, $B_{\nu'}(T)$ is the Plank function evaluated at the emitted frequency, $\nu'$, and $\kappa_\nu$ is the dust opacity.  Using the range 
of 
observed Bolocam flux densities (4.0 - 6.8 mJy), assuming a redshift of 2.1, and applying the dust 
cross section of $\kappa_\nu$ = 12.4 cm$^2$/g \citep{ossenkopf94} most relevant for high mass star formation (high gas density and thin ice mantle model), leads to dust masses of 
3 - 5 $\times$ 10$^8$ M$_\odot$ (2 - 3 $\times$ 10$^8$ M$_\odot$) for a dust temperature of 40 K (50 K).  Blindly applying a dust-to-gas mass ratio of 1/100 implies gas masses of 
3 - 5 $\times$ 10$^{10}$ M$_\odot$.  These gas masses are comparable to those from a recent sample of 8 submillimeter galaxies of \cite{genzel04} and
\cite{neri03} which yield median molecular gas masses (from CO emission) of 2.2 $\times$ 10$^{10}$ M$_\odot$ and 2.8 $\times$ 10$^{10}$ M$_\odot$, with median dynamical masses of 1.1 $\times$ 
10$^{11}$ M$_\odot$ and 6.2 $\times$ 10$^{10}$ M$_\odot$ (assuming the most probable inclination angle of sin $i = 2/\pi$), respectively.  These gas mass estimates are uncertain to at least a factor 
of a few due to 1) the Bolocam flux density bias, which causes the measured flux densities to be overestimated by 10 - 30\%, 2) our assumed values of $\kappa$ 
and T, which may vary by a factor of a few and $\pm$ 20 K, respectively, and 3) increasing our assumed redshift of 2.1 (the mean of the 5 Chapman spectroscopic 
redshifts of the Bolocam galaxies) to $z$ = 2.4 {\it increases} our calculated gas mass by 30\%.
Nevertheless, taking these factor into account still imply that a considerable fraction of the mass could already by in stars and substantial gas remains for star formation.  This 
major epoch of galaxy formation at approximately $z$ = 2 is consistent with the conclusions of \cite{fontana04} from spectral fitting of a sample of 500 elliptical galaxies at 0.2 $\le$ $z$ $\le$ 
2.5 that approximately 35\% of elliptical galaxy stellar mass was assembled by $z$ = 2 and approximately 80\% by $z$ = 1.

\section{Conclusions}

        We have obtained 350 $\mu$m SHARC II observations toward galaxy candidates from the Bolocam Lockman Hole survey.  The
Lockman Hole has rich, deep, multiwavelength observations enabling detailed studies of galaxies.  The 350 $\mu$m photometry is
near the peaks of the SEDs of galaxies with characteristic temperatures of $\sim$ 50 K and redshifts of $z$ $\sim$ 2 to 3.  They therefore
enable measurements of luminosities and estimates of temperatures and photometric redshifts without interpolating over the
peak of the FIR thermal SEDs.  Seven galaxies detected at 1.1 mm with Bolocam were detected at 350 $\mu$m, two of which have two
350 $\mu$m counterparts; these were combined with two 350 $\mu$m detections from the survey of \cite{kovacs05}, bringing the 
total number of Bolocam galaxies detected with SHARC II to nine.  Two additional galaxies not associated with the Bolocam sources
were also detected.  The SHARC II detections range in significance from 3.0 $\sigma$ to 6.8 $\sigma$, with flux densities
ranging from 14 mJy to 64 mJy.

        We combined our observations with 850 $\mu$m and 1.2 mm photometry from the literature to fit the
submillimeter/millimeter-wave spectra to thermal dust models.  We found that two models with significantly different dust
temperatures (40 K and 68 K) and spectral indices $\beta$ (1.6 and 1.8, respectively) yielded similar quality fits owing to the
degeneracy in T and $\beta$, rendering them indistinguishable without better SED sampling.  However, there is little
consequence of the degeneracy to the derived luminosities, photometric redshifts, and dust masses within the statistical
uncertainties.  Five of the galaxies have spectroscopic redshifts in the literature, with four ranging from $z$ = 2.1 to 3.0 and
one at $z$ = 0.689.  The four high-z galaxies have luminosities of (1.0 - 1.2) $\times$ 10$^{13}$ L$_\odot$, while the $z$ = 0.689 galaxy is
best fit by a 20 K, $\beta$ = 1.0, spectrum with a much lower luminosity:  1.3 $\times$ 10$^{11}$ L$_\odot$.  (Given the source confusion in the optical and radio, along with consistent photometric 
redshifts, we suggest that the $z$ = 0.689 spectroscopic redshift of Bolocam source 17 may be a misidentification.)  The characteristic dust masses
for the four high-$z$ spectroscopic galaxies are 4 $\times$ 10$^8$ M$_\odot$, implying gas masses of 4 $\times$ 10$^{10}$
M$_\odot$.  The dominant uncertainties in this estimation are the dust opacity and the gas-to-dust conversion factor, which
make the estimation uncertain to a factor of a few.  Assuming a Salpeter IMF and that the submillimeter emission derives 
completely from star formation yields stellar masses of 10$^{10}$ to a few times 10$^{11}$ M$_\odot$, broadly consistent with the
stellar content of modern-day elliptical galaxies.

        The photometric redshifts of the full sample of seven galaxies span the range of $z$ = 2.0 to $z$ = 4.3, with statistical
uncertainties of $\Delta z$ = 0.3 to 0.6 (1 $\sigma$).  Photometric redshifts utilizing composite radio/FIR spectra representative
of local star-forming ULIRGs yields systematically higher redshifts, on the order of $\Delta z$ = 1.  For the four galaxies
with optical spectroscopic redshifts the anomolously high redshifts arise from systematically low 1.4 GHz observed flux
densities.  The discrepancy could arise from small number statistics, inverse-Compton losses of high energy cosmic rays off the CMB, 
heating by radio-quiet AGN, or suppressed synchrotron emission from supernova remnants in the unusually luminous galaxies.  For comparison,
photometric redshifts derived using only Spitzer MIPS and IRAC data points yielded slightly more precise and accurate
redshifts than the submillimeter/millimeter-wave data alone, with discriminatory power between heating by AGN and star
formation (albeit with limited bolometric luminosity constraints).

\acknowledgments

We thank Attila Kov\'{a}cs for providing us with SHARC II detections of Bolocam sources 5 and 8.   
We also acknowledge the support of the CSO director and staff, the support of Kathy Deniston, and
helpful conversations with Steven Eales.  
This work was supported in part by NSF grants AST-0098737, AST-9980846, and AST-0206158 and
PPARC grants PPA/Y/S/2000/00101 and PPA/G/O/2002/00015.  G.\ T.\ L.\ acknowledges NASA for GSRP Fellowship NGT5-50384, D.\ J.\ Haig and D.\ Dowell for their 
assistance during the SHARC II observing runs, and the entire Bolocam instrument team.

\clearpage

\appendix
\section{Multiwavelength Coverage}
\label{section:coverage}

Observations at both longer and shorter wavelengths are necessary to characterize high redshift submillimeter galaxies.  Radio 
identifications provide precise astrometry which allow optical and infrared counterparts to be identified.  The radio observations detect the synchrotron 
emission (presumably from high energy cosmic rays associated with supernova explosions, see review by Condon 1992), which is known to be correlated to the far-infrared luminosity (and is thus 
correlated to the Bolocam flux density, which traces the dust emission that is responsible for the infrared luminosity).  The Bolocam and radio points may then be used to obtain temperature and 
redshift estimates.  Multiple submillimeter wavelengths are essential in confirming the Bolocam sources and allow for a photometric redshift measurement that is independent 
of the radio-to-far-IR correlation (i.e. using only the submillimeter portion of the spectrum for an assumed temperature).  Furthermore, a third photometric redshift estimate using the 1.6 $\mu$m 
bump in the stellar emission spectrum is possible using the near- and mid-infrared portion of the spectrum.  In this appendix we briefly discuss the multiwavelength coverage (submillimeter, 
radio, 
infrared, optical, and X-ray) of the Lockman Hole 
and the coincident detections of each survey.  Figure \ref{fig:circles} presents an overview of the Bolocam detections, including all of the submillimeter, radio, and infrared 
counterparts.

\subsection{Submillimeter Surveys}

Four submillimeter / millimeter surveys of the Lockman Hole region are considered.  Three of the surveys 
(Bolocam, SCUBA, MAMBO) represent blank-field surveys, while the SHARC II survey covers only the 17 Bolocam sources.  The surveys include:

	1) Entire Bolocam Lockman Hole survey at 1.1 mm \citep{laurent05}.  The survey covers 324 arcmin$^2$ down to an RMS of
        1.4 mJy/beam and includes 17 source candidates ($\ge$ 3 $\sigma$) ranging in flux densities from 4.0 to 6.8 mJy.  Extensive simulations and jackknife 
	tests yielded 6 expected false detections.  The Bolocam beam size is 31$\arcsec$ and observations were taken without chopping while scanning at 60$\arcsec$/s.  The Bolocam survey 
includes 
the entire 
	regions surveyed by the published 8 mJy 850 $\mu$m JCMT SCUBA \citep{scott02} and 1.2 mm IRAM MAMBO \citep{greve04} surveys.  Using a model SED 
	based on nearby, dusty, star-forming galaxies (see \S\ \ref{section:photometric_redshifts}) gives relative flux densities of 1\,:\,2.0\,:\,0.8 
	and relative RMSs of 1\,:\,0.9\,:\,0.6$-$1.4 for the Bolocam, SCUBA, and MAMBO surveys, respectively, for a galaxy redshift of $z=2.4$ (with the 
	range given for MAMBO due to nonuniform noise).

	2) JCMT SCUBA 8 mJy survey \citep{scott02}.  The survey covers 122 arcmin$^2$ (in the Lockman Hole) to an RMS of 2.5
        mJy/beam with a 14$\arcsec$ beam, and includes 36 source candidates ($\ge$ 3 $\sigma$).  
	Five sources of the \cite{scott02} catalog (LE850.9, 10, 11, 15, 20) were retracted by \cite{ivison02} on the basis of large $\sigma_{850\mu m}$ values (and lack
        of radio identifications).  The SCUBA survey detects 6 of the 8 Bolocam detections in the overlap region between the surveys (Bolocam 5, 8, 13, 14, 16, and 17).

	3) 1.2 mm IRAM MAMBO survey \citep{greve04}.  The survey covers 197 arcmin$^2$ (in the Lockman Hole) to an RMS of 0.6-1.5 mJy/beam, 
	with a 10.7$\arcsec$ beam. 
	The survey includes 23 sources ($\ge$
        3 $\sigma$).  The MAMBO survey detects 7 of the 11 Bolocam detections in the overlap region between the surveys (Bolocam 1, 4, 5, 8, 14, 16, and 17).

	4) 350 $\mu$m SHARC II observations of the Lockman Hole (this work -- See \S\ \ref{section:sharc}).  The instrument beam is 9$\arcsec$ and
        the observations were taken with a Lissajous scan pattern with uniform coverage regions of 95$\arcsec$ $\times$ 18$\arcsec$ to an RMS of 6-18
        mJy/beam (with deeper integrations taken for fainter sources).  A total of 14 sources were detected \citep[3 by][]{kovacs05}, 11 of which 
	coincide with 9 Bolocam sources.  (An additional 3 SHARC II sources were detected outside of the Bolocam positional error circles.)

	To summarize, the majority of the Bolocam sources within the regions covered by other submillimeter surveys were detected (SCUBA detects 6 of 8 
	Bolocam detections, MAMBO detects 7 of 11, and SHARC II detects 9 of 17).  Seven of the Bolocam source candidates were not detected by any other submillimeter
	survey (Bolocam 7, 9, 10, 11, 12, 13, and 15, each of which was detected at $<$ 3.5 $\sigma$), which is consistent with the 6 false detections (Poisson distributed) 
	expected from simulations.  These results point to the robustness of the remaining Bolocam galaxy candidates.

\subsection{Radio Detections}

	Three independent VLA radio surveys were used to identify counterparts to the Bolocam sources.  In addition, this work re-analyzes the 
\cite{ivison02} data set to obtain a complete source catalog over the entire Bolocam source catalog.  The radio surveys include:

	\noindent {\it Published Surveys:}

        1) VLA radio source list at 1.4 GHz of the SCUBA 8 mJy survey sources \citep{ivison02}.  The Lockman East radio survey reached an average noise level of 4.8 $\mu$Jy 
	beam$^{-1}$ with a 1.3$\arcsec$ FWHM circular beam.  Coincident sources (and $5 \sigma$ upper limits) are given for each of the 21 
	brighter ($> 3 \sigma$) SCUBA sources, with 20 positive detections over 14 of the sources.  Five of the 21 SCUBA sources (LE850.9, 10, 11, 15, 20) are ignored on the basis 
	of large $\sigma_{850\mu m}$ values (and lack of radio identifications).  Of the 8 Bolocam detections covered by this VLA survey, 5 (Bolocam 5, 8, 14, 16, 17) coincide with 7 
	\cite{ivison02} sources.  \cite{ivison02} also report continuum data at 4.9 GHz (using the VLA in the C configuration), with a 
	resulting noise level of $\sim$ 11 $\mu$Jy beam$^{-1}$.  Positive detections exist for 8 of the SCUBA galaxies, 4 of which are coincident with Bolocam sources (Bolocam 5, 
	14, 16, 17).

        2) A source list from the 6 cm VLA survey of the Lockman Hole \citep{ciliegi03}.  The survey has a noise level of $\sim 11 \mu$Jy and covers the northeast quadrant of the 
	Bolocam survey, including 7 of the Bolocam sources.  A total of 63 sources were detected at $>$ 4.5 $\sigma$, with a radio counterpart for only one of the Bolocam sources 
	(Bolocam 8).

	\noindent {\it Unpublished Surveys:}

       	3) A source list of 20 cm VLA observations was obtained from \cite{yun05}.  The observations
        were broken into four fields, with the northeastern section being deeper (10 $\mu$Jy/beam RMS) than the other three fields
        (15 $\mu$Jy/beam).  The observations cover the entire Bolocam field, with a total of 640 sources $> 3.5 \sigma$.  The source list contains 
	counterparts for 10 of the Bolocam sources (Bolocam 1, 2, 3, 5, 6, 8, 9, 13, 14, 17).

        4) Re-analysis of the VLA map from \cite{ivison02} by \cite{biggs06}.  The survey reached a median noise level in the locality of the 17 Bolocam sources of 4.4 $\mu$Jy beam$^{-1}$.  
	Radio sources were extracted from the VLA map within 30$\arcsec$ from the Bolocam source positions.  Counterparts for 15 of the 17 Bolocam sources were detected (all except Bolocam 
sources 10 and 12).  
	Seven of the detections are coincident with the \cite{ivison02} analysis (Bolocam 5, 8, 14, 16, and 17), with flux densities in good agreement (within the stated uncertainties).
	
	In total, 15 of the 17 Bolocam galaxy candidates are found to have at least one radio counterpart (all except Bolocam 10 and 12).  As only the VLA observations from \cite{yun05}
	and \cite{ciliegi03} include an entire source catalog, we are limited to estimating accidental detection rates for these two surveys.  The accidental 
	detection rate is defined as the Poisson likelihood that one or more of these known radio sources, randomly distributed, fall within the 2 $\sigma$ positional confidence region of 
	the Bolocam beam (see \S\ \ref{section:positionalerrors}).  The rates for the \cite{yun05} and \cite{ciliegi03} surveys are 17-24\% and 8-11\%, 
	respectively.  (The accidental radio detection rate is dominated by the Yun et al.\ 2005 survey due to its greater relative depth.)

\subsection{Infrared Detections}

	Both published and unpublished {\it Spitzer} observations were used to obtain infrared counterparts to the Bolocam / radio sources.  They 
include:

       1) {\it Spitzer} postage stamps and an extracted source list (this work) with the IRAC 
        (3.6, 4.5, 5.8, and 8.0 $\mu$m) and MIPS (24 $\mu$m) instruments were obtained for 14 of the 17 Bolocam source candidates for
        areas within a 16$\arcsec$ radius of the Bolocam beam centers.  (Note that this radius is smaller than the 2 $\sigma$
        positional errors of Bolocam.)  Bolocam.LE.1100.2, 6, and 12 fall outside of the region observed with {\it Spitzer}.
        Multiple {\it Spitzer} sources were detected within each of the 14 Bolocam positional error circles.

        2) {\it Spitzer} observations of SCUBA/VLA selected sources \citep{egami04} with IRAC (3.6, 4.5, 5.8, 8.0 $\mu$m) and MIPS
        (24 $\mu$m) in a 5' x 5' area of the Lockman Hole region.  The MIPS 24 $\mu$m field of view is 5.4' x 5.4'.  The total
        integration time was 300 seconds per pixel, with a 3 $\sigma$ detection limit of 120 $\mu$Jy.  The observed region contains the following 
	{\it Spitzer}, VLA radio, SCUBA, and Bolocam correlations:

\begin{itemize}

\item The field imaged by IRAC and MIPS contains 10 SCUBA sources of the \cite{scott02} catalog (LE850.1, 4, 7, 8, 10, 14, 18, 23, 24, 35).
        
\item \cite{ivison02} detected radio counterparts for 5 (LE850.1, 7, 8, 14, 18) of the 7 secure ($\ge$ 3.5 $\sigma$) SCUBA sources at high 
significance and a radio counterpart for LE850.4 with lower significance.  

\item \cite{egami04} finds an additional radio counterpart to LE850.4 upon re-examination of the VLA map.  

\item Of these 7 SCUBA sources with a total of 9 radio components (LE850.1, 4, 7, 8a, 8b, 14a, 14b, 18, 35), all are found to have {\it Spitzer} counterparts.
        
\item {\it Spitzer} counterparts are found only for the sources with radio detections.  

\item Four of the 9 {\it Spitzer} detections (LE850.1, 8a, 8b, 8c) fall within 2 of the Bolocam positional error circles (Bolocam.LE.1100.14 and 17).

\end{itemize}

\subsection{Optical Detections}

Optical counterparts to the Bolocam / radio sources were obtained from three deep surveys, including:

	1) SDSS u, g, r, i, z photometry was obtained from the public SDSS data release 3 (DR3) website\footnote{http://cas.sdss.org/dr3/en/tools/search/rect.asp}, including 7,148 objects 
	in the Lockman Hole region.  The 61 objects falling within the Bolocam positional error circles include 43 galaxies (extended) and 18 stars (point-like).  

        2) A list of objects with Johnson R-band (SUBARU) photometry was obtained from \cite{yun05} for each of the \cite{yun05} VLA 20 cm radio sources.  The optical 
	catalog consists of 1,031 objects down to a limiting magnitude of R = 24.5.  29 sources were found within the positional error circles of the 10 Bolocam sources covered by the 
	catalog.  As no radio sources were detected by \cite{yun05} for Bolocam sources 4, 7, 10, 11, 12, 15, and 16, no optical coverage was available for these sources.

        3) Optical Johnson R-band (SUBARU) photometry was also obtained from \cite{ivison05} for most of the Lockman Hole region (with no coverage for 
	Bolocam sources 6, 9, and 11).  50,297 sources were detected down to a limiting magnitude of R = 27.  950 sources lie within the positional error circles of the 14 Bolocam sources 
	covered by the survey.  

%

\subsection{X-ray Detections}
\label{subsection:xray}

	Finally, two X-ray surveys were used to identify Bolocam sources with a possible AGN contribution.  The X-ray surveys include: 

	1) The ROSAT Ultra Deep Survey \citep{lehmann01} source list of the Lockman Hole was cross referenced with the positions of the 17 Bolocam galaxy candidates.  The survey 
	reached a flux density level of 1.2 $\times$ 10$^{-15}$ erg cm$^{-2}$ s$^{-1}$ in the 0.5-2.0 keV energy band, detecting 94 X-ray sources.  Two of the sources coincide with the Bolocam 
	positions, coincident with the eastern radio source of Bolocam 8 and in-between the two radio sources associated with Bolocam 17.  Despite the positional offset, \cite{lehmann01}
	associate the X-ray emission coincident with Bolocam source 17 with the northwestern radio (and optical) source.  An additional X-ray detection exists of SHARC II source 3 
	(in the field of Bolocam source 9), although the SHARC II and X-ray positions lie well outside the positional error circle of the Bolocam source.

	2) XMM-Newton observations of the Lockman Hole ($\simeq$ 100 ksec) were taken during the Performance Verification phase of the instrument \citep{mainieri02}, yielding 98 
	sources with more than 70 net counts (flux density limit of 1.6 $\times$ 10$^{-15}$ erg cm$^{-2}$ s$^{-1}$ in the 0.5-7 keV band).  Cross referencing the source list with the 17 Bolocam 
	sources yields a coincidence list identical to that of the ROSAT survey.

	3) A deep, (1 Msec) unpublished \citep{ivison05} XMM-Newton survey of the majority of the Bolocam good coverage region (Bolocam 2, 6, 11, and 12 are not covered) confirms the above results with only one identification, 
	coincident with Bolocam source 17.

\begin{rotate}
\setlength{\tabcolsep}{0.5mm}
\begin{deluxetable}{ccccccccccccccccccccc}
\tabletypesize{\tiny}
\tablecaption{Summary of Multiwavelength Detections of Bolocam Galaxy Candidates}
\tablewidth{0pt}
\tablehead{
\colhead{Bolocam} & \colhead{Bolocam} & \colhead{SHARC II} & \colhead{SCUBA} & \colhead{MAMBO} & \colhead{Yun} & \colhead{Yun (deep)} & \colhead{Biggs}	
& \colhead{Ivison} & \colhead{Ivison} & \colhead{Ciliegi} & \colhead{{\it Spitzer}} & \colhead{{\it Spitzer}} & \colhead{{\it Spitzer}}  & \colhead{{\it Spitzer}} & \colhead{{\it Spitzer}} 
& \colhead{SDSS\tablenotemark{a}} & \colhead{Yun\tablenotemark{b}} & \colhead{Ivison} & \colhead{Notes} \\
\colhead{Number} & \colhead{1.1 mm} & \colhead{350 $\mu$m} & \colhead{850 $\mu$m} & \colhead{1.2 mm} & \colhead{20 cm} & \colhead{20 cm} &\colhead{20 cm} 	
& \colhead{20 cm} & \colhead{6 cm} & \colhead{6 cm} & \colhead{3.6 $\mu$m} & \colhead{4.5 $\mu$m} & \colhead{5.8 $\mu$m} & \colhead{8.0 $\mu$m} & \colhead{24 $\mu$m} 
& \colhead{u,g,r,i,z} & \colhead{R} & \colhead{R} & \colhead{} \\
\colhead{} & \colhead{mJy} & \colhead{mJy} & \colhead{mJy} & \colhead{mJy} & \colhead{$\mu$Jy} & \colhead{$\mu$Jy} &\colhead{$\mu$Jy} 	
& \colhead{$\mu$Jy} & \colhead{$\mu$Jy} & \colhead{$\mu$Jy} & \colhead{mJy} & \colhead{mJy} & \colhead{mJy} & \colhead{mJy} & \colhead{mJy} 
& \colhead{} & \colhead{} & \colhead{} & \colhead{}
}
\startdata
1 &6.8$\pm$1.4&38.0$\pm$9.3&X           &5.7$\pm$1.0&-         &60$\pm$13	  &52$\pm$5.7	     &X		&X	&-	&1(4)&1(4)&-(-)&-(-)&1(-)&-(1,1)  		 	&-&1(48)&S radio source \\
" &"	      &"           &X           &"          &-         &-                 &53$\pm$5.6        &X		&X	&-	&1   &1   &-   &1   &-   &-  		 	&X&-    &N radio source \\
2 &6.5$\pm$1.4&20.9$\pm$5.2&X           &X          &373$\pm$58&X     	          &507$\pm$9.0	     &X		&X	&X	&X   &X   &X   &X   &X   &1,0(0,1)		 	&1&1(74)&N radio source\\
" &"          &"           &X           &X          &-         &X		  &90$\pm$9.7	     &X		&X	&X	&X   &X   &X   &X   &X   &-       		 	&X&1    &S radio source\\
3 &6.0$\pm$1.4&15.1$\pm$3.8&X\tablenotemark{c}&-    &241$\pm$39&X                 &289$\pm$9.0       &X		&X	&X	&1(10)&1(4)&1(1)&1(-)&1(2)&0,1(0,2)\tablenotemark{c}&2\tablenotemark{d}&1(35)&NE SHARC II source\\
" &"          &14.0$\pm$3.5&X\tablenotemark{c}&-    &84$\pm$20 &X                 &96$\pm$7.8        &X		&X	&X	&1   &1   &-   &-   &1   &-       		 	&-&1    &SW SHARC II source\\
4 &5.2$\pm$1.4&-           &-           &3.2$\pm$0.7&-         &X                 &20$\pm$6.7        &X		&X	&X	&X\tablenotemark{c}(14)&X\tablenotemark{c}(13)&X\tablenotemark{c}(1)&X\tablenotemark{c}(1)&X\tablenotemark{c}(3)&1,0(3,2)&X&1(41)&\\
5 &5.1$\pm$1.3&35.6$\pm$9.7&11$\pm$2.6  &2.9$\pm$0.7\tablenotemark{c}&X&51$\pm$13 &51$\pm$9.3        &54$\pm$14	&60$\pm$35&-	&1(10)&1(6)&1(1)&1(-)&1(2)&-(2,0)                  	&3\tablenotemark{d}&1(47)&E radio source\\
" &"          &-           &-           &"          &X         &59$\pm$13         &42$\pm$6.5        &X		&X	&-	&1    &1   &1   &1   &1   &1,0                     	&3&1    &S radio source\\
" &"          &-           &-           &3.1$\pm$0.7&X         &-                 &27$\pm$5.9        &X		&X	&-	&X    &X   &X   &X   &X   &-                       	&X&-    &W radio source\\
" &"          &-           &-           &-          &X         &-                 &29$\pm$7.8        &X		&X	&-	&X    &X   &X   &X   &X   &-                       	&X&-    &N radio source\\
6 &5.0$\pm$1.5&27.6$\pm$6.9&X           &X          &60$\pm$17 &X                 &55$\pm$9.9        &X		&X	&X	&X    &X   &X   &X   &X   &1,0(4,0)                	&-&X    &E SHARC II, E radio source\\
" &"          &"           &X           &X          &105$\pm$22&X                 &138$\pm$11        &X		&X	&X	&X    &X   &X   &X   &X   &-                       	&-&X    &E SHARC II, W radio source\\
" &"          &63.6$\pm$15.8&X          &X          &611$\pm$93&X                 &939$\pm$16        &X		&X	&X	&X    &X   &X   &X   &X   &1,0                     	&1&X   &W SHARC II source\\
7 &4.9$\pm$1.5&-           &X           &X          &-         &X                 &51$\pm$23         &X		&X	&X	&-(6) &-(5)&-(1)&-(-)&-(2)&-(-)                    	&X&-(71)&\\
8 &4.8$\pm$1.3&34.5$\pm$9.3&10.9$\pm$2.4&4.8$\pm$0.6&X         &-		  &27$\pm$7.6        &29$\pm$11	&-	&-	&1(13)&1(13)&1(-)&-(-)&1(6)&-(2,1)		 	&X&1(83)&S radio source\\
" &"          &"           &"           &"          &X         &46$\pm$12         &28$\pm$6.7        &24$\pm$9	&-	&-	&1   &1   &-   &1   &1   &-                        	&-(2)&1 &N radio source\\
" &"          &"           &-           &-          &X         &136$\pm$23        &157$\pm$8.0       &X		&X	&58$\pm$11&X   &X   &X   &X   &X   &0,1                      	&3   &1 &E radio source\\
9 &4.8$\pm$1.5&-\tablenotemark{c}&X     &X          &60$\pm$18 &X		  &68$\pm$11         &X		&X	&X	&1(13)&1(14)&1(-)&1(-)&1(5)&-(1,0)                 	&2&X    &\\
10&4.7$\pm$1.4&-           &-           &-          &-         &X                 &-                 &X		&X	&X	&-(6)&-(6)&-(3)&-(1)&-(1)&-(2,1)                   	&X&-(60)&\\
11&4.6$\pm$1.5&-           &X           &X          &-         &X                 &116$\pm$33        &X		&X	&X	&1(6)&1(4)&-(-)&-(-)&-(-)&1,0(3,2)                 	&X&X    &\\
12&4.6$\pm$1.4&-\tablenotemark{c}&X     &X          &-         &X                 &-                 &X		&X	&X	&X   &X   &X   &X   &X   &-(3,0)                   	&X&-(66)\tablenotemark{c}&\\
13&4.5$\pm$1.4&-           &-\tablenotemark{c}&-    &63$\pm$18 &X                 &63$\pm$7.2        &X		&X	&X	&-(7)&-(7)&-(1)&-(-)&-(2)&1,0(3,0)                 	&3\tablenotemark{d}&1(58)&NE radio source\\
" &"          &-           &-           &-          &-         &X                 &27$\pm$7.9        &X		&X	&X	&1   &1   &1   &-   &1   &0,1                      	&X&1    &E radio source\\
" &"          &-           &-           &-          &-         &X                 &30$\pm$7.6        &X		&X	&X	&-   &-   &-   &-   &-   &-                        	&X&-    &SE radio source\\
14&4.4$\pm$1.3&24.1$\pm$6.0&10.5$\pm$1.6\tablenotemark{c}&3.4$\pm$0.6&76$\pm$19&78$\pm$15&73$\pm$7.3 &73$\pm$10	&56$\pm$37&-	&X\tablenotemark{c}(7)&X\tablenotemark{c}(3)&X\tablenotemark{c}(-)&X\tablenotemark{c}(-)&X\tablenotemark{c}(1)&-(1,1)  &-&1(83)&S radio source\\
" &"          &-           &-           &-          &-         &X                 &34$\pm$7.2        &X		&X	&-	&1   &1   &1   &1   &1   &-                        	&X&1    &W radio source\\
" &"          &-           &-           &-          &-         &45$\pm$12         &26$\pm$8.4        &X		&X	&-	&X\tablenotemark{c}&X\tablenotemark{c}&X\tablenotemark{c}&X\tablenotemark{c}&X\tablenotemark{c}&-&1&1&NE radio source\\
" &"          &-           &-           &-          &-         &-                 &26$\pm$8.3        &X		&X	&-	&-   &-   &-   &-   &-   &1,0                      	&X&1 &SW radio source\\
15&4.4$\pm$1.4&-           &X           &-          &-         &X                 &50$\pm$7.7        &X		&X	&-	&1(13)&1(12)&1(3)&1(2)&1(2)&1,0(5,0)               	&X&1(60)&\\
16&4.1$\pm$1.4&44.0$\pm$9.0&6.1$\pm$1.8 &2.8$\pm$0.5&X         &-                 &49$\pm$8.5        &41$\pm$12	&32$\pm$22&-	&1(9) &1(7) &1(-)&1(-)&1(-)&-(1,1)                 	&X&1(112)&\\
17&4.0$\pm$1.3&15.5$\pm$3.9&5.1$\pm$1.3 &2.4$\pm$0.6&-         &-                 &20$\pm$6.6        &22$\pm$11	&-	&-	&1(4) &1(5) &1(1)&1(0)&1(2)&0,1(3,2)               	&1&1(91)&NW radio source\\
" &"          &"           &"           &"          &-         &54$\pm$13         &58$\pm$9.4        &58$\pm$12	&57$\pm$32&-	&1   &1   &1   &1   &1   &-                        	&3(4)&1 &SE radio source\\
\enddata
\tablenotetext{a}{The two listed numbers correspond to the SDSS galaxy and star classifications, respectively.}
\tablenotetext{b}{The Optical R band source list \citep{yun05} includes only sources within 3$\arcsec$ of the corresponding radio source.}
\tablenotetext{c}{See source description in \S\ \ref{section:summary}}
\tablenotetext{d}{The multiple \cite{yun05} R-band optical sources coincident with the radio source are close in both position 
($\le$ 1$\arcsec$) and flux ($\le$ 0.2 mag), and are likely due to a lack of merging close sources in the catalog.}
\tablecomments{An (X) corresponds to Bolocam sources not covered by the survey / source list.  Items in parentheses correspond to the detection of sources within the Bolocam 
positional error circle, but not likely associated with the source.}
\label{table:detections}
\end{deluxetable}
\setlength{\tabcolsep}{2mm}
\end{rotate}

\section{Previous Redshift Estimates of Bolocam Galaxies}
\label{section:previousredshift}

Several of the Bolocam galaxies with radio counterparts have had their redshifts determined with either photometric or 
spectroscopic techniques.  We briefly discuss the results of each of these redshift surveys in this section.  The spectroscopic redshifts will form the basis of comparison to the photometric 
redshift techniques discussed in \S\ \ref{section:photometric_redshifts}.  The previous photometric redshifts also prove useful in considering the merits of including 350 $\mu$m SHARC 
II observations when estimating redshifts of submillimeter galaxies.

	1) \cite{chapman05} obtained spectroscopic redshifts for a sample of 73 submillimeter galaxies, the largest set of spectroscopic redshifts to date.  The sources are a subset 
	of 150 SCUBA/JCMT sources detected at 850 $\mu$m over seven separate fields.  Requiring accurate positions for spectroscopic identification, 104 of the 
	SCUBA galaxies have radio 
	identifications from deep VLA radio maps at 1.4 GHz.  Of these, a subset of 98 sources were observed with the Low Resolution Imaging Spectrograph (LRIS) on the Keck I telescope, 
	resulting in the spectroscopic identification of 73 of these galaxies.  Twelve of the sources correspond to the SCUBA sources in the Lockman Hole, 5 of which a coincident with 
	Bolocam galaxy candidates (Bolocam 5, 8, 14, 16 and 17).  \cite{chapman05} find redshifts of 2.611, 3.036, 2.148, 2.142, 0.686 for the five Bolocam galaxies, respectively.  
	While spectra for each of the Bolocam galaxies are not included among the sample spectra, \cite{chapman05} comment that all of the above detections have multiple-line 
	identifications and can therefore be considered robust.  The subset of \cite{chapman05} redshifts with Bolocam detections ($z = 2.1 \pm 0.9$) is consistent with the overall 
	redshift distribution of the entire sample (with a median redshift of 2.2).  
  
	2) \cite{aretxaga03} implement a Monte Carlo photometric redshift technique, using existing far-IR - radio multiwavelength data for 77 sources first identified at 850 $\mu$m 
	(SCUBA) or 1.2 mm (MAMBO).  The technique produces the redshift probability distribution for an individual galaxy by choosing an evolutionary model and generating a catalog of 
	luminosities and redshifts based on template SEDs from local starbursts, ULIRGs, and AGN.  Thirteen of the sources correspond to SCUBA sources in the 
	Lockman Hole, including  
	5 Bolocam galaxies (Bolocam 5, 8, 14, 16 and 17 -- the same subset with Chapman et al.\ 2005 spectroscopic redshifts).  Five different evolutionary models were used, with the 
	results listed in Table \ref{table:previousredshifts}.
	Given the stated uncertainties in the photometric redshifts, the \cite{aretxaga03} results are in general agreement with the spectroscopic measurements.  Note that the photometric redshift for 
	Bolocam source 17 is consistent with both the modified blackbody and near IR stellar bump models (see \S\ \ref{subsection:bolocam17}), further supporting a possible misidentification by 
	\cite{chapman05}.
	
\setlength{\tabcolsep}{1mm}
\begin{deluxetable}{ccccccc}
\tabletypesize{\tiny}
\tablecaption{Previous Spectroscopic and Photometric Redshifts of Bolocam Galaxies}
\tablewidth{0pt}
\tablehead{
\colhead{Bolocam} & \colhead{Chapman} & 	\colhead{Aretxaga} & 				
	\colhead{Lehmann}  &	\colhead{Oyabu}    &    \colhead{Egami}\\
\colhead{Number} & \colhead{et al.\ (2005)} & 	\colhead{et al.\ (2003)} & 			
	\colhead{et al.\ (2001)} &	\colhead{et al.\ (2005)} &    \colhead{et al.\ (2004)}\\
\colhead{} & \colhead{$z_{\mathrm{spec}}$} & 	\colhead{$z_{\mathrm{phot}}^{\mathrm{le2}}$} & 	
	\colhead{$z_{\mathrm{spec}}$} &		\colhead{$z_{\mathrm{spec}}$}   &   \colhead{$z_{\mathrm{phot}}$}
}
\startdata
5	&2.611(NE)	&2.7$^{+1.8}_{-0.7}$(NE)	&		&		&\\
6	&		&				&		&0.362(W)	&\\
8	&3.036(S)	&5.8$^{+0.2}_{-1.7}$		&		&1.110(E)	&\\
14	&2.148(S)	&2.6$^{+0.4}_{-0.5}$(S)		&		&		&2.6(S)\\
16	&2.142		&3.3$^{+1.1}_{-1.3}$		&		&		&\\
17	&0.689(SE)	&3.7$^{+1.5}_{-0.7}$		&0.974(NW)	&		&$\sim$3(SE), 0.9\\
\enddata
\label{table:previousredshifts}
\end{deluxetable}
\setlength{\tabcolsep}{2mm}

	3) The ROSAT Ultra Deep Survey \citep{lehmann01} of the Lockman Hole reached a flux density level of 1.2 $\times$ 10$^{-15}$ erg cm$^{-2}$ s$^{-1}$ in the 0.5-2.0 keV energy 
	band detecting 94 X-ray sources.  Spectroscopic identifications of 90\% of the sources based on highly accurate positions (2$\arcsec$ FWHM) with the High Resolution Imager (HRI) 
and 
	Keck R-band images were performed with the LRIS spectrometer on the Keck telescopes, yielding a single spectroscopic redshift for the northwest optical component of Bolocam source 17.  
	Multiple lines were detected (Mg II, [Ne V], and [O II]) giving a redshift of 0.974 (cf.\ \S\ \ref{subsection:bolocam17}).

	4) \cite{oyabu05} spectroscopically identified 29 of the 44 far-infrared sources (with accurate 1.4 GHz VLA radio positions) detected in the 0.9 deg$^2$ {\it ISO} survey 
	of the Lockman Hole.  Optical identifications of the radio sources were performed in both the I- and R-bands (using the University of Hawaii 88$\arcsec$ and Subaru telescopes), 
with optical 
	spectroscopy performed with both the Keck II and WIYN telescopes.  Five of the {\it ISO} sources fall in the region covered by Bolocam, with 2 sources coinciding with Bolocam sources (6 and 
	8).  \cite{oyabu05} find a redshift of 0.362 for the most westerly radio source of Bolocam 6 and a redshift of 1.110 for eastern radio source of Bolocam 8.  

	5) With {\it Spitzer} IRAC and MIPS, \cite{egami04} positively detect all 9 of the radio sources associated with 7 SCUBA galaxies, including 4 detections 
	of two Bolocam sources (14 and 17).  The photometric redshifts using only the near- and mid-infrared points are 2.6 for Bolocam 14, and $\sim$ 3 and 0.9 for the two radio sources 
	near Bolocam 17.

\end{document}